\pdfoutput=1
\newcommand{\be}{\begin{equation}}
\newcommand{\ee}{\end{equation}}
\newcommand{\bea}{\begin{eqnarray}}
\newcommand{\eea}{\end{eqnarray}}
\newcommand{\ba}[1]{\begin{array}{#1}}
\newcommand{\ea}{\end{array}}
\documentclass[preprint,secnumarabic,amssymb,nobibnotes,nofootinbib,aps,pra,showpacs]{revtex4}
\usepackage{epsfig}
\usepackage{amsmath}
\usepackage{amssymb}
\usepackage{mathrsfs}
\usepackage{color}
\usepackage{times}
\usepackage{mathtools}
\usepackage{mathrsfs}
\usepackage{graphicx}
\usepackage{epstopdf}

\begin{document}

\title{Parametric oscillations in a dissipative bosonic Josephson junction}
\author{Abhik Kumar Saha$^1$, Deb Shankar Ray$^2$, and Bimalendu Deb$^{1}${\footnote{Corresponding author, e-mail: msbd@iacs.res.in}}}
\affiliation{$^1$School of Physical Sciences, 
$^2$School of Chemical Sciences,   Indian Association
for the Cultivation of Science,
Jadavpur, Kolkata 700032, India.}
\def\zbf#1{{\bf {#1}}}
\def\bfm#1{\mbox{\boldmath $#1$}}
\def\hf{\frac{1}{2}}
\begin{abstract}
We study the dynamics of a nonlinear dissipative bosonic Josephson junction (BJJ) with a time-dependent sinusoidal perturbation in interaction term. We demonstrate parametric resonance where the system undergoes sustained periodic oscillations even in the presence of dissipation. This happens when the frequency of the perturbation is close to twice the frequency of the unperturbed Josephson oscillations and the strength of perturbation exceeds a critical threshold. We have formulated the threshold conditions for parametric oscillations. To explore the nature of the oscillations, we carry out a multiple time scale analysis of the stability boundaries in terms of the V-shaped Arnold's tongue in the parameter space. Full numerical simulations have been performed for the zero-, running- and $\pi$-phase modes of nonlinear Josephson effect. Our results demonstrate that in $\pi$-phase mode, the system is capable of making a transition from regular parametric to chaotic parametric oscillations as one crosses the stability boundary. Also, the phase difference undergoes phase slip before executing sustained parametric oscillations.   
\end{abstract}

\maketitle

\section{Introduction}
Ultracold atomic gases provide a unique platform for simulating and exploring many condensed-matter phenomena \cite{Bloch:RMP,Bloch:nat,Zhai:rep}, including Josephson effect \cite{Joseph} which is an unambiguous manifestation of macroscopic quantum coherence existing between two superfluids or superconductors. In this context a paradigmatic model is the BJJ in a double-well (DW) potential \cite{shenoy}. Josephson oscillations in BJJ have been reported in number of experiments \cite{joseph:exp,joseph:exp1,joseph:exp2,albiez}. Though most of the experimental and theoretical works on ultracold atomic Josephson effect deal with non-dissipative Josephson effect and related phenomena such as macroscopic quantum self-trapping (MQST) \cite{shenoy,joseph:exp2}, in recent times dissipative Josephson effect has attracted a considerable amount of research interest \cite{joseph:dam1,pigneur,dam:3}. A few recent experiments have reported damping of Josephson oscillations \cite{joseph:dam,pigneur1}. The question we address here is  how to suppress or mitigate the damping of BJJ in order to get sustained periodic oscillations. In this work, we show that it is indeed possible to overcome the effect of damping by add a periodic time-dependent perturbation in a suitable parameter of the system.

Parametric oscillations of a physical system can be achieved by periodically varying one of its parameters to modulate the natural frequency of the oscillator \cite{landau,arnold}. Such phenomena are ubiquitous in classical physical systems such as the vertically driven pendulum \cite{feynman}, Paul ion trap \cite{paul}, and aspects of some models of the universe \cite{kofman}. In the context of ultracold quantum gases, there have been some studies related to the parametric resonances such as Faraday patterns \cite{faraday:1,faraday:2,faraday:3,faraday:4,faraday:5,faraday:6,faraday:7}, barrier resonance \cite{barrier:1}, bright and vortex solitons \cite{soliton:1,soliton:2,soliton:3}, self-damping at zero temperature \cite{ref:1}, condensates in oscillating DW potential \cite{osc:1,osc:2}, quasi-particle creation and thermalization \cite{quasi:1},  periodic modulation of interatomic interaction in a ring trap \cite{quasi:3}. Parametric resonances also arise when an optical lattice is shaken \cite{dalfovo} and in Kelvin waves of a quantized vortex line in trapped Bose-Einstein condensates \cite{kelvin}.  However, to the best of our knowledge, the parametric oscillations in dissipative BJJ have not been studied so far.

The basic paradigm of such phenomena is described by Mathieu-Hill equation $\ddot{x}+a(t)x=0$, where $a(t)=a(t+T)$ with $T=2\pi/\omega$, the time period of the parameter $a(t)$. If $T$ or $2T$ matches the integral multiples of the natural time period ($\tau=2\pi/\omega_{H}$, where $\omega_{H}$ is the frequency of the unperturbed system) of oscillation, parametric resonance occurs causing instability in the dynamics. The main feature of the parametric oscillation is that a nonlinear dynamical system remains in an oscillatory state in absence of any  additive forcing term  when the characteristic time period of the unperturbed system or any of its multiples matches the time period of the parameter.

The aim of the present work is to examine the dynamic response of a dissipative BJJ to a time-periodic variation of a suitable parameter in the interaction term. We observe that a sinusoidal time-dependent perturbation in interaction  can give rise to sustained periodic oscillations  in a dissipative BJJ under certain specific conditions beyond a critical threshold that are compatible with parametric resonances. We investigate the stable and unstable regions by a multiple time scale analysis for a range of frequencies and amplitudes of the time-dependent part of the interaction. The dependence is portrayed in a typical ``Arnold tongue" in a graph of perturbation amplitude vs. perturbation frequency. A theoretical analysis is carried out to elucidate the characteristic parametric oscillations for the three well-known phase modes of nonlinear BJJ, namely zero-, running- and $\pi$-phase modes. Full numerical simulations demonstrate that in $\pi$-phase mode one observes transition from regular parametric oscillation to chaotic parametric oscillation as one crosses the stability boundary. The dynamics of phase difference exhibits an interesting phase slip before the system undergoes sustained regular oscillations.

The paper is organized in the following way. In Sec. \ref{2}, we analyze the theoretical method of formulating parametric oscillations in dissipative BJJ. The conditions for critical threshold, parametric resonance and the dynamical solutions in zero-, running-, and $\pi$-phase modes are derived in Sec. \ref{3}.  Sec. \ref{4} is devoted to a multiple time scale analysis to identify the stable and unstable regions. In Sec. \ref{5}, we present and discuss our results on numerical simulations to corroborate the theoretical scheme. The paper is concluded in Sec. \ref{6}.

\section{The dissipative BJJ Model}\label{2}
For a Bose-Einstein condensate (BEC) trapped in a DW potential, we define the normalized atom number imbalance 
\begin{equation}
w=\frac{N_{L}-N_{R}}{N_{L}+N_{R}}
\end{equation}
with $N_{L,(R)}$ the number of atoms in left (right) well. The conjugate variable is relative phase defined by
\begin{equation}
 \phi=\phi_{L}-\phi_{R}
\end{equation}
with $\phi_{L,(R)}$ phase of the atoms in left (right) well.

Usually, Josephson oscillations in a DW potential is nondissipative meaning that the dynamics of the atom number imbalance and relative phase remains undamped over time \cite{albiez,shenoy,joseph:exp1}. However, in recent times several studies have reported dissipative BJJ which is analogous to a pendulum with friction \cite{pigneur1,pigneur,marino,dam:3}. The governing equations of a dissipative BJJ are 
\begin{equation}
\dot{w}(t)=-\frac{2J}{\hbar}{\sqrt{1-w^2(t)}}\sin\phi(t)-\frac{\eta}{N}\dot{\phi}(t)
\end{equation}
\begin{equation}
\dot{\phi}(t)=\frac{2J}{\hbar}\left[\Lambda_{0} w(t)+\frac{w(t)}{\sqrt{1-w^2(t)}}\cos\phi(t)\right]
\end{equation}
where $\eta$ is the viscosity due to dissipation, $J$ is the tunneling energy, $N$ is the total atom number $N=N_{L}+N_{R}$ and $\Lambda_{0}=\frac{NU_{0}}{2J}$ characterizes the many-body interaction parameter with $U_{0}$ being the on-site mean two-body interaction energy. We consider a trap potential which allows harmonic oscillations along radial directions ($x$- and $y$- axes) and a symmetric DW potential along $z$- axis. The form of the DW potential \cite{abhik:jpb} is $V_{dw} (z)= \frac{1}{2}\chi_{0}^2(z^2-b^2)^2$, where $z={\pm b}$ are the two minimum points where the DW potential vanishes and the barrier height is $V_0=\frac{1}{2}\chi_{0}^2 b^4$. We now add a sinusoidal perturbation term to $V_{0}$ changing $V_{0}$ to $V_{t}$, that is, barrier height becomes oscillating. As a result, the time-dependent DW potential is 
\begin{equation}
 V_{dw}(z,t)=\frac{1}{2}\chi^2(z^2-b^2)^2
\end{equation}
where $\chi^2=\chi_{0}^2+\chi_{1}^2\sin(\omega_{p}t)$, where $\chi_{0}$ is the unperturbed part and $\omega_{p}$ is the frequency of the input time-dependent sinusoidal perturbation which makes the barrier oscillate. Experimentally, a sinusoidally oscillating $U$ can be obtained by making the barrier height of a DW trap to oscillate with small frequency and amplitude by modulating the trapping frequencies or amplitudes of external fields (lasers in case of an optical trap, or radio-frequency fields in case of a magnetic trap). The presence of an oscillating barrier causes the BEC to fluctuate around the ground state. As a result,  the on-site interaction energy $U$ becomes time-dependent. Assuming that $\chi_{1}<<\chi_{0}$, under tight-binding or two-mode approximation of the DW potential, $U=U_{0}(1+\zeta\sin\omega_{p}t)$, where $\zeta=\frac{\chi_{1}^2}{4\chi_{0}^2}$ is dimensionless and small quantity.  That a sinusoidal oscillation of  barrier can lead to a temporal oscillations in $U$ is discussed in the Appendix. Under this condition, $\Lambda$ becomes time-dependent.  
\begin{equation}
\Lambda(t)=\Lambda_{0}+h\sin(\omega_{p}t)  
\end{equation}
where $h=\Lambda_{0}\zeta$ is the amplitude of the input time-dependent sinusoidal perturbation and $\Lambda_{0}$ is the unperturbed part. In this context, it is to be noted that the temporal oscillation of $\Lambda$  has been previously  used in Ref. \cite{salasnich} to discuss  parametric resonance in a non-dissipative BJJ.

Now,  let us consider that this dynamical system has a steady state at ($w_s,\phi_{s}$). Now putting the value of $\Lambda$(t) in Eq. (4) and linearizing the system around the steady state ($w_s,\phi_{s}$) with $w=w_{s}+\delta w$ and $\phi=\phi_{s}+\delta\phi$, we obtain
\begin{equation}
 \delta\dot{w}=-\frac{2J}{\hbar}{\sqrt{1-w_{s}^2}}\cos\phi_{s}\delta\phi+\frac{2J}{\hbar}\frac{w_{s}\delta w}{\sqrt{1-w_{s}^2}}\sin\phi_{s}  -\frac{\eta}{N}\delta\dot{\phi}
\end{equation}
\begin{equation}
 \delta\dot{\phi}=\frac{2J}{\hbar}\left[\Lambda(t)\delta w-\frac{w_{s}\sin\phi_{s}}{\sqrt{1-w_{s}^2}}\delta\phi+\frac{\cos\phi_{s}}{(1-w_{s}^2)^{\frac{3}{2}}}\delta w \right]
\end{equation}
By differentiating Eqs. (7) and (8) with respect to time, we write
\begin{equation}
\delta\ddot{w}=-\frac{2J}{\hbar}{\sqrt{1-w_{s}^2}}\cos\phi_{s} \delta\dot{\phi}+\frac{2J}{\hbar}\frac{w_{s} \delta\dot {w}}{\sqrt{1-w_{s}^2}}\sin\phi_{s} -\frac{\eta}{N}\delta\ddot{\phi} 
\end{equation}
\begin{equation}
\delta\ddot{\phi}=\frac{2J}{\hbar}\left[\Lambda(t)\delta\dot {w}-\frac{w_{s}\sin\phi_{s}}{\sqrt{1-w_{s}^2}}\delta\dot{\phi}+\frac{\cos\phi_{s}}{(1-w_{s}^2)^{\frac{3}{2}}}\delta\dot {w}+\dot{\Lambda}(t)\delta w \right] 
\end{equation}
Substitution of Eqs. (8) and (10) in Eq. (9) results in a damped parametric oscillator \cite{ghosh} governed by
\begin{equation}
\delta\ddot{w}+\left[\kappa+\frac{2J\eta}{\hbar N}h\sin\omega_{p}t\right]\delta\dot{w}+\left[\omega_{J}^2+\frac{4J^2}{\hbar^2}h\sin\omega_{p}t+\frac{2J\eta}{\hbar N}h\omega_{p}\cos\omega_{p}t\right]\delta w=0
\end{equation}
where $\omega_{J}^2=\frac{4J^2}{\hbar^2}[\Lambda_{0}+1]$ and $\kappa=\frac{2J\eta}{\hbar N}[\Lambda_{0}+1]$. Here we choose the value of steady state $(w_{s},\phi_{s})=(0,0)$ which is one of the stable steady states as in BJJ \cite{shenoy}. We now rescale time $t$ as a dimensionless time $t2J/\hbar\rightarrow t$ and perturbation frequency $\omega_{p}$ as a dimensionless perturbation frequency $\frac{\omega_{p}\hbar}{2J}\rightarrow \omega_{p}$. Eq. (11) then becomes
\begin{equation}
\delta\ddot{w}+\left[\kappa+\frac{\eta}{N}h\sin\omega_{p}t\right]\delta\dot{w}+\left[\omega_{J}^2+h\sin\omega_{p}t+\frac{\eta}{N}h\omega_{p}\cos\omega_{p}t\right]\delta w=0 
\end{equation}
where $\omega_{J}$ and $\kappa$ are dimensionless. Eq. (12) describes an oscillator in which Josephson frequency $\omega_{J}$ and damping $\kappa$ are modulated by a perturbation term with sinusoidal time dependence.

\section{Parametric resonance in dissipative BJJ model; theoretical considerations}\label{3}
We now look for the analytical solutions of damped parametric oscillator described by Eq. (12). The main characteristic of the damped parametric oscillator is that it is capable of sustained periodic oscillations, say, at $\omega$, i.e., the frequency of the perturbed dynamics. To show this, we assume a solution 
\begin{eqnarray}
 \delta w(t)=A\cos(\omega t+\theta)
\end{eqnarray}
where $A$ and $\theta$ are the amplitude and phase of the solution signal wave, respectively. Expanding $\sin\omega_p t$ and $\cos\omega_p t$ in Eq. (12) in terms of exponentials and substituting Eq. (13) and neglecting non-synchronous terms oscillating at $\omega_p+\omega$, we are led to the following equation
\begin{eqnarray}
 \left[\omega_{J}^2-\omega^2+i\kappa\omega\right]e^{i(\omega t+\theta)}+\left[-\frac{ih}{2}+\frac{\eta\omega_p h}{2N}-\frac{\eta\omega h}{2N}\right]e^{i(\omega_p t-\omega t-\theta)}=0
\end{eqnarray}
From Eq. (14), it follows that sustained oscillation is possible if 
\begin{eqnarray}
 \omega_{p}=2\omega 
\end{eqnarray}
and equating real and imaginary parts of the Eq. (14), we get,  
$\omega_{J}^2-\omega^2+\frac{\eta\omega_{p}h}{2N}-\frac{\eta h \omega}{2N}=0$, $\theta=0,m\pi$, where $m$ is an integer and 
\begin{eqnarray}
h=2\kappa\omega
\end{eqnarray}
In other words, when the perturbation frequency $\omega_p$ is twice the oscillation frequency $\omega$ and phase $\theta=0$ or $m\pi$, the strength of perturbation $h$ must satisfy Eq. (16). The last condition is the threshold $h_{t}$ for oscillations, since it assumes a perturbation strength $h$ necessary to overcome the mean losses ($\kappa$) at the oscillation threshold. This implies that the system undergoes spontaneous oscillation at a higher strength of $h$ for a frequency $\omega=\omega_{p}/2$ as a result of continuous transfer of energy from the source at $\omega_p$ to the system mode at $\omega_{p}/2$ when the threshold $h_{t}$ is crossed.
The presence of $\phi$ in $\Lambda_{0}$ in Eq. (16) by virtue of the relation $\kappa=\frac{\eta}{N}[1+\Lambda_{0}]$ makes the dynamics of parametric oscillation dependent on phase and in what follows we demonstrate the role of this phase in phase slip in dissipative BJJ, particularly in $\pi$-phase mode. Maintaining this condition on phase $\theta$ by adjusting the perturbation parameters opens a new perspective for applications of BJJ.

Secondly, it is important to emphasize that the condition of parametric resonance implies $\omega_{p}=2\omega\simeq2\omega_{J}\simeq2\sqrt{1+\Lambda_{0}}$; $\omega_{p}$ is thus not the frequency of any external drive as used in usual parametric resonance phenomena. $\omega_{p}$ is a characteristic of the interaction term itself because of the presence of $\Lambda_{0}$. We now look for the solutions of the linearized dissipative BJJ in absence of perturbation in the zero-, running-, $\pi$-phase modes.

\subsection{Zero-phase mode}
This mode describes the tunneling dynamics when the average of the population imbalance and phase across the junction is zero. In zero-phase mode if the input time-dependent perturbation terms are zero (i.e., $h=0$) then the Eq. (12) reduces to the 
\begin{eqnarray}
\delta\ddot{w}+\kappa\delta\dot{w}+\omega_{J}^2\delta w=0  
\end{eqnarray}
which is the equation of motion of a damped harmonic oscillator. The zero-phase mode frequency is 
\begin{eqnarray}
 \omega_{0}=\sqrt{1+\Lambda_{0}-\frac{\eta^2}{4N^2}(1+\Lambda_{0})^2}
\end{eqnarray}
and the characteristic decay time
\begin{eqnarray}
 \tau_{0}=\frac{2N}{\eta(1+\Lambda_{0})}
\end{eqnarray}
In the absence of damping and perturbation terms Eq. (12) reduces to that of the well-known BJJ, used for analysis of stability for Josephson oscillations and MQST.

\subsection{Running-phase mode}
One of the main features in BJJ is MQST that can be achieved when the tunneling is strongly suppressed and the particles remain mostly trapped in one of the wells, as a result the average of population imbalance remains non-zero. In order to reach MQST, one has to increase the initial population imbalance $w(0)$ above a critical value for fixed $\Lambda_{0}$ or alternatively increase $\Lambda_{0}$ by changing the interaction parameters keeping $w(0)$ fixed \cite{shenoy}.  There are two different types of MQST depending on the time evolution of $\phi$. If it evolves unbounded increasing (or decreasing) always in time, it is called running-phase mode. In this mode, the expressions for the characteristic mode frequency and the decay time remain same as in the zero-phase mode. The only difference is that the value of $\Lambda_{0}$ should be above of the critical value. As a result, the characteristic frequency $\omega_{0}$ becomes non-zero as $\Lambda_{0}$ is increased above the critical value when damping is present \cite{marino}. Unlike the case when there is no damping $(\eta=0)$, $\omega_{0}$ dips to zero as $\Lambda_{0}$ is increased above the critical value \cite{shenoy}.

\subsection{$\pi$-phase mode}
Apart from the zero- and running-phase modes, BJJ has another important class of tunneling dynamics in which the system evolves with a time-averaged value $\pi$ of the phase difference. Since this dissipative BJJ model relies on the classical analogy with the momentum shortened pendulum \cite{marino}, it allows the pendulum to perform small and large amplitude $\pi$ oscillations with average value of $w$ being zero around an unstable equilibrium. This is similar to a vertically upward oriented pendulum. The dynamics of the population imbalance changes to a macroscopically self-trapped mode with non-zero average of $w$ if $\Lambda_{0}$ exceeds a critical value. This is closely analogous to the rotation of an inverted pendulum with a closed loop trajectory.

Linearizing Eqs. (3) and (4) around the steady state values $w_{s}=0$, $\phi_{s}=\pi$, we get
\begin{eqnarray}
\delta\ddot{w}-\kappa\delta\dot{w}+\omega_{J}^2\delta w=0  
\end{eqnarray}
where $\omega_{J}^2=1-\Lambda_{0}$ and $\kappa=\frac{\eta}{N}(1-\Lambda_{0})$ and $\Lambda_{0}<1$. As a result, $\pi$-phase mode frequency becomes 
\begin{eqnarray}
\omega_{\pi}=\sqrt{1-\Lambda_{0}-\frac{\eta^2}{4N^2}(1-\Lambda_{0})^2}
\end{eqnarray}
and the characteristic decay time
\begin{eqnarray}
 \tau_{\pi}=\frac{2N}{\eta(1-\Lambda_{0})}
\end{eqnarray}
In the $\pi$-phase mode, we get two types of MQST characterized by the time-averaged value of the population imbalance $w<w_{s}$ and $w>w_{s}$ with $w_{s}$ being the steady state value of $w$ at which symmetry breaking occurs. $w_{s}$ defined as $w_{s}=\sqrt{1-\frac{1}{\Lambda_{0}^2}}$ \cite{shenoy}. When the system in the MQST state and if their is no perturbation then linearizing the Eqs. (3) and (4) around the steady state values $w_{s}=\sqrt{1-1/\Lambda_{0}^2}$, $\phi_{s}=\pi$, we get 
\begin{eqnarray}
\delta\ddot{w}-\frac{\eta\Lambda_{0}}{N}(\Lambda_{0}^2-1)\delta\dot{w}+(\Lambda_{0}^2-1)\delta w=0  
\end{eqnarray}
The frequency of the self-trapped dynamics
\begin{eqnarray}
 \omega_{ST}=\sqrt{\Lambda_{0}^2-1-\frac{\eta^2\Lambda_{0}^2}{4N^2}(\Lambda_{0}^2-1)^2}
\end{eqnarray}
with characteristic decay time
\begin{eqnarray}
 \tau_{ST}=\frac{2N}{\eta\Lambda_{0}(\Lambda_{0}^2-1)}
\end{eqnarray}
and maintaining the condition $\Lambda_{0}>1$.   

Before concluding this section, we mention that the parametric oscillation frequency $\omega$ is determined by the perturbation frequency $\omega_{p}$ through the Eq. (16). However, the characteristic frequencies stated in the above depend only on the system parameters while parametric oscillation frequency $\omega$ exists for every $\omega_{p}$ and the oscillations occur when the threshold condition is crossed.

\section{A multiple time scale analysis of stable and unstable regions}\label{4}
We now return to Eq. (12) and resort to a multiple time scale analysis. The main idea behind this analysis is to locate the region where the system looses its stability and search for characteristic solutions. To do this, we first rewrite dynamical Eq. (12) for $\delta w(t)$ in a modified time scale $\tau=\omega_p t$ as follows
\begin{eqnarray}
\delta\ddot{w}+\epsilon\rho\left[1+c\sin\tau\right]\delta\dot{w}+\left[\gamma+\epsilon\sin\tau+\epsilon\rho c\cos\tau\right]\delta w=0 
\end{eqnarray}
where, $\epsilon=\frac{h}{\omega_p^2}$, $\rho=\frac{\omega_p\kappa}{h}$, $c=\frac{\eta h}{N\kappa}$, $\gamma=\frac{\omega_J^2}{\omega_p^2}$.

Our approach is based on two time scale expansion method for Eq. (26) for small values of $\epsilon$. Eq. (26) constitutes two time scales, the time scale $\xi=\tau$ of the periodic motion itself and a slower time scale $\sigma=\epsilon\tau$ which represents the approach to the periodic motion. Now, if we expand $\delta w(\xi,\sigma)$ in a power series in $\epsilon$ as
\begin{eqnarray}
\delta w(\xi,\sigma)=\delta w_{0}(\xi,\sigma)+\epsilon\delta w_{1}(\xi,\sigma)+\epsilon^2\delta w_{2}(\xi,\sigma) 
\end{eqnarray}
and connect it to Eq. (26), it is clear that the resulting equation which can be solved by order of $\epsilon$. So, zero order equation of $\epsilon$ can be represented as 
\begin{eqnarray}
\frac{\partial^2}{\partial\xi^2}(\delta w_{0})+\gamma(\delta w_{0})=0
\end{eqnarray}
which gives a solution of a simple harmonic oscillator with frequency $\sqrt{\gamma}$.
\begin{eqnarray}
\delta w_{0}=A(\sigma)\cos(\sqrt{\gamma}\xi)+B(\sigma)\sin(\sqrt{\gamma}\xi)
\end{eqnarray}
and by first order equation of $\epsilon$, we obtain
\begin{eqnarray}
\frac{\partial^2}{\partial\xi^2}(\delta w_{1})+\gamma(\delta w_{1}) &=& -2\frac{\partial^2}{\partial\sigma\partial\xi}(\delta w_{0}) -\rho[1+c\sin\xi]\frac{\partial}{\partial\xi}(\delta w_{0})\nonumber\\ &-&\sin\xi(\delta w_{0}) - \rho c \cos\xi(\delta w_{0}) 
\end{eqnarray}
Therefore substituting Eq. (29) into Eq. (30) and then using standard trigonometric identities, we arrive at the following equation
\begin{eqnarray}
\frac{\partial^2}{\partial\xi^2}(\delta w_{1})+\gamma(\delta w_{1}) &=& \left[2\sqrt{\gamma}\frac{dA}{d\sigma}+\sqrt{\gamma}A\rho\right]\sin\sqrt{\gamma}\xi -\left[2\sqrt{\gamma}\frac{dB}{d\sigma}+\sqrt{\gamma}B\rho\right]\cos\sqrt{\gamma}\xi\nonumber\\ &+&\left[\frac{\rho cA\sqrt{\gamma}}{2}-\frac{B}{2}\right][\cos(1-\sqrt{\gamma})\xi-\cos(1+\sqrt{\gamma})\xi]\nonumber\\ &-& \left[\frac{\rho cB\sqrt{\gamma}}{2}+\frac{A}{2}\right][\sin(1+\sqrt{\gamma})\xi+\sin(1-\sqrt{\gamma})\xi]\nonumber\\ &-& \frac{\rho cA}{2}[\cos(1+\sqrt{\gamma})\xi+\cos(1-\sqrt{\gamma})\xi]\nonumber\\ &-& \frac{\rho cB}{2}[\sin(1+\sqrt{\gamma})\xi-\sin(1-\sqrt{\gamma})\xi]
\end{eqnarray}
Now, if we choose $\gamma=\frac{1}{4}$ then Eq. (31) becomes
\begin{eqnarray}
\frac{\partial^2}{\partial\xi^2}(\delta w_{1})+\gamma(\delta w_{1})&=&\left[\frac{dA}{d\sigma}+\frac{\rho A}{2}+\frac{\rho c B}{4}-\frac{A}{2}\right]\sin\frac{\xi}{2}\nonumber\\&-& \left[\frac{dB}{d\sigma}+\frac{\rho B}{2}+\frac{\rho c A}{4}+\frac{B}{2}\right]\cos\frac{\xi}{2}\nonumber\\ &-&\left[\frac{3\rho c B}{4}+\frac{A}{2}\right]\sin\frac{3\xi}{2}+\left[-\frac{3\rho c A}{4}+\frac{B}{2}\right]\cos\frac{3\xi}{2}
\end{eqnarray}   
To avoid secular terms, we set the coefficients of $\sin\frac{\xi}{2}$ and $\cos\frac{\xi}{2}$
equal to zero so that we have

\begin{eqnarray} 
\begin{pmatrix}
\frac{dA}{d\sigma}  \\
\frac{dB}{d\sigma}  
\end{pmatrix}
=
\begin{pmatrix}
-\frac{\rho}{2}+\frac{1}{2} & -\frac{\rho c}{4} \\
-\frac{\rho c}{4} & -\frac{\rho}{2}-\frac{1}{2} 
\end{pmatrix}
\begin{pmatrix}
A \\
B 
\end{pmatrix}
\end{eqnarray}
By solving the above equation, we get condition for which $A$ and $B$ have exponential growth. This instability arises because of $\gamma=\frac{1}{4}$ and corresponds to a 2:1 subharmonic resonance in which the perturbation frequency ($\omega_p$) is twice the Josephson frequency ($\omega_J$). Expanding $\gamma$ in a power series of $\epsilon$, one obtains
\begin{eqnarray}
 \gamma=\frac{1}{4}+\epsilon\gamma_{1}+\epsilon^2\gamma_{2}+......
\end{eqnarray}
Repeating the same calculation with $\gamma$ as stated in Eq. (34), we get additional terms in Eq. (33) as follows:
\begin{eqnarray} 
\begin{pmatrix}
\frac{dA}{d\sigma}  \\
\frac{dB}{d\sigma}  
\end{pmatrix}
=
\begin{pmatrix}
-\frac{\rho}{2}+\frac{1}{2} & -\frac{\rho c}{4}+\gamma_{1} \\
-\frac{\rho c}{4}-\gamma_{1} & -\frac{\rho}{2}-\frac{1}{2} 
\end{pmatrix}
\begin{pmatrix}
A \\
B 
\end{pmatrix}
\end{eqnarray}
The above equation can be solved by assuming a solution in the form $A(\sigma)=A_{0}exp(\sigma\lambda)$, $B(\sigma)=B_{0}exp(\sigma\lambda)$. For nontrivial constants $A_{0}$ and $B_{0}$, the following condition must hold:
\[
\begin{vmatrix}
-\frac{\rho}{2}+\frac{1}{2}-\lambda & -\frac{\rho c}{4}+\gamma_{1} \\
-\frac{\rho c}{4}-\gamma_{1} & -\frac{\rho}{2}-\frac{1}{2}-\lambda
\end{vmatrix}
=
0
\]
The eigenvalues $\lambda_{\pm}$ are given by $\lambda_{\pm}=-\frac{\rho}{2}\pm \sqrt{\frac{\rho^2c^2}{16}-\gamma_{1}^2+\frac{1}{4}}$. For the transitions between stable and unstable regions we set $\lambda_{\pm}=0$ giving the value for $\gamma_{1}=\sqrt{\frac{\rho^2c^2}{16}-\frac{\rho^2}{4}+\frac{1}{4}}$. This condition gives the two transition curves emerging from $\gamma=\frac{1}{4}$ in the form of a V-shaped profile known as Arnold's tongue in the $\epsilon$-$\gamma$ plane. $\gamma$ is therefore modified upto first order as 
\begin{eqnarray}
\gamma=\frac{1}{4}\pm\epsilon\sqrt{\frac{\rho^2c^2}{16}-\frac{\rho^2}{4}+\frac{1}{4}}
\end{eqnarray}

\section{Numerical results and discussions}\label{5}
In this section, we present the numerical results in zero-, running-, and $\pi$-phase modes. One object is to analyse the effect of sinusoidal periodic perturbation on the dissipative BJJ above and below of the critical threshold. 

\subsection{Parametric resonances in zero-phase mode}
\begin{figure}[h]
    \centering
    \begin{minipage}{.32\textwidth}
        \centering
        \includegraphics[width=2.26in, height=1.8in]{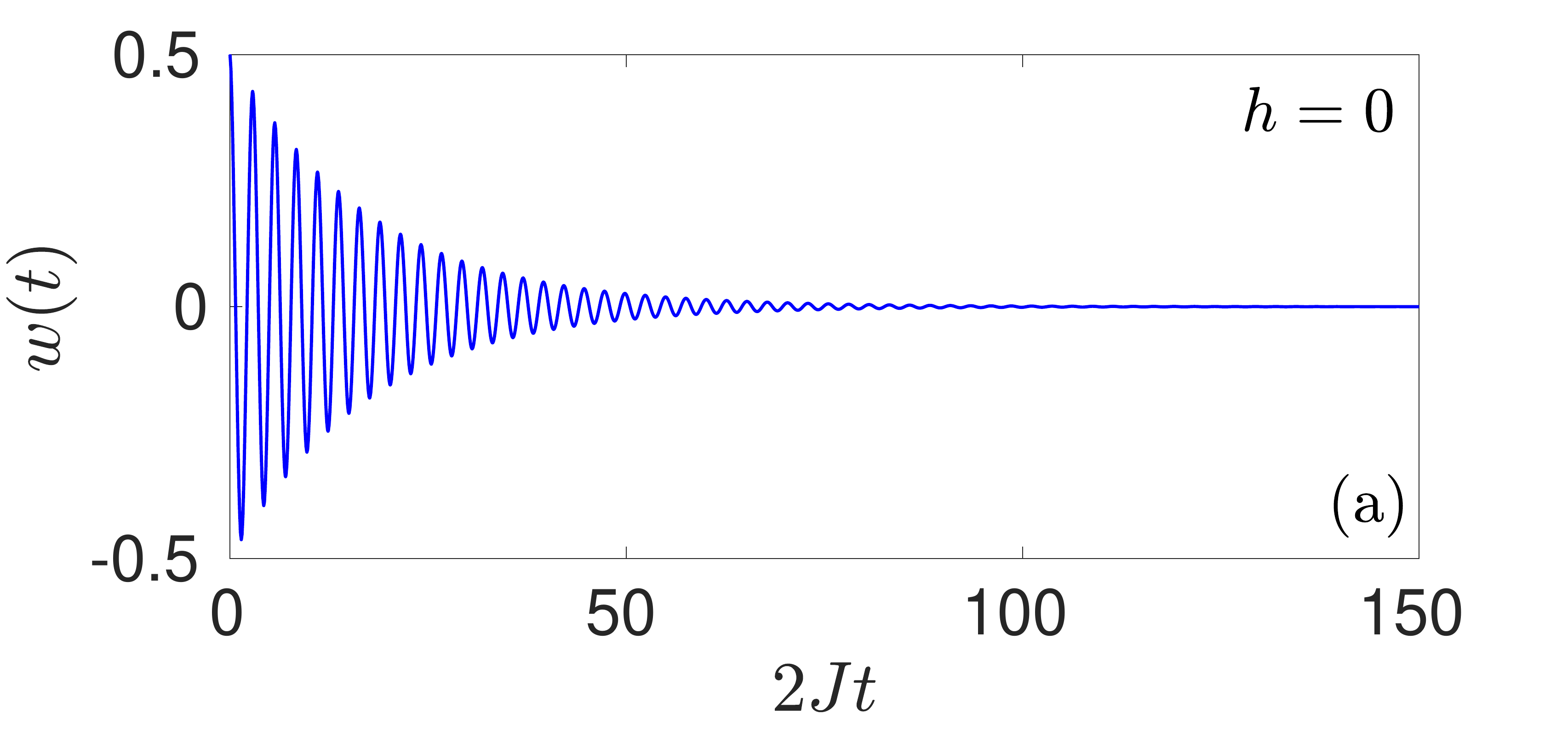}
    \end{minipage}
    \begin{minipage}{0.32\textwidth}
        \centering
        \includegraphics[width=2.26in, height=1.8in]{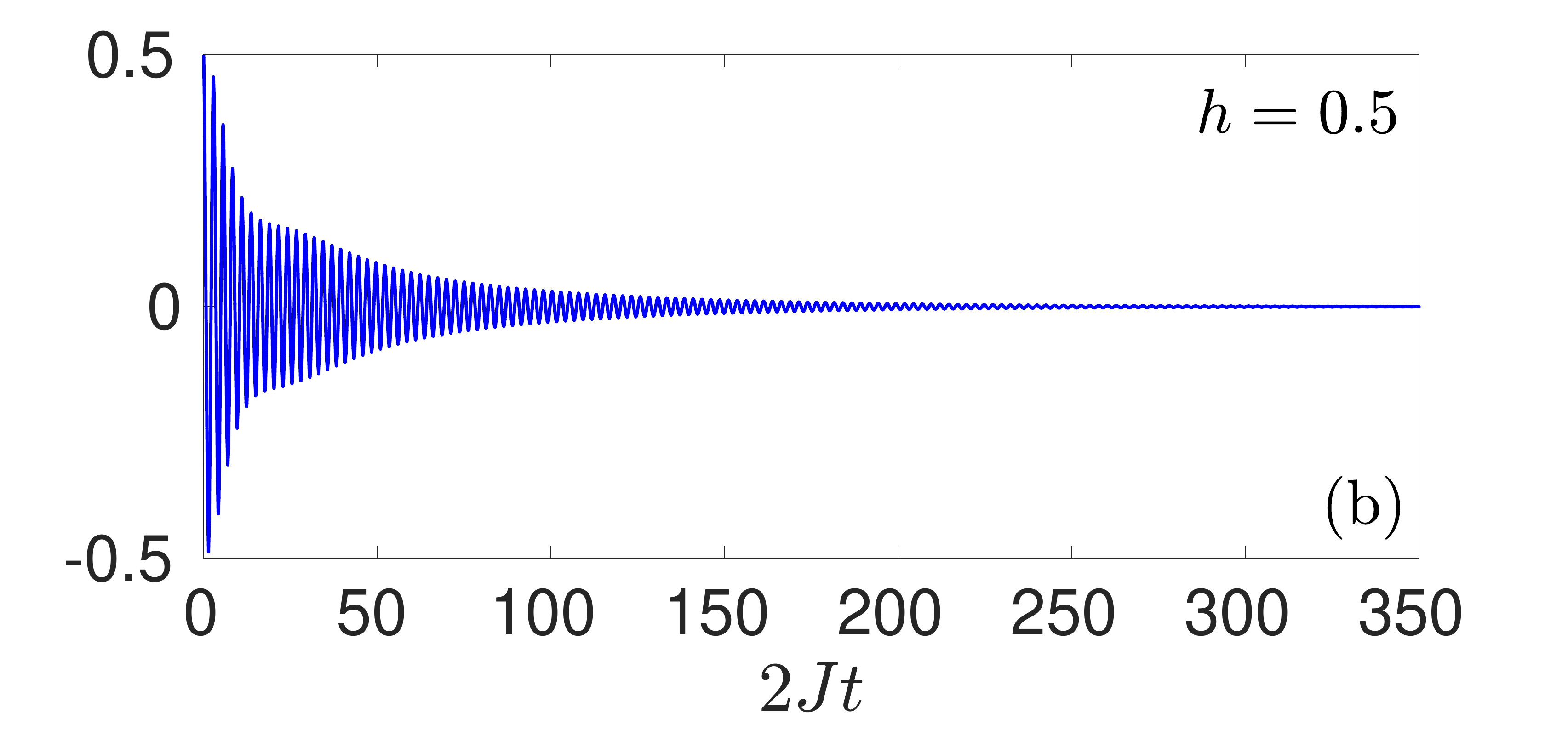}
    \end{minipage}
        \begin{minipage}{0.32\textwidth}
        \centering     
        \includegraphics[width=2.26in, height=1.8in]{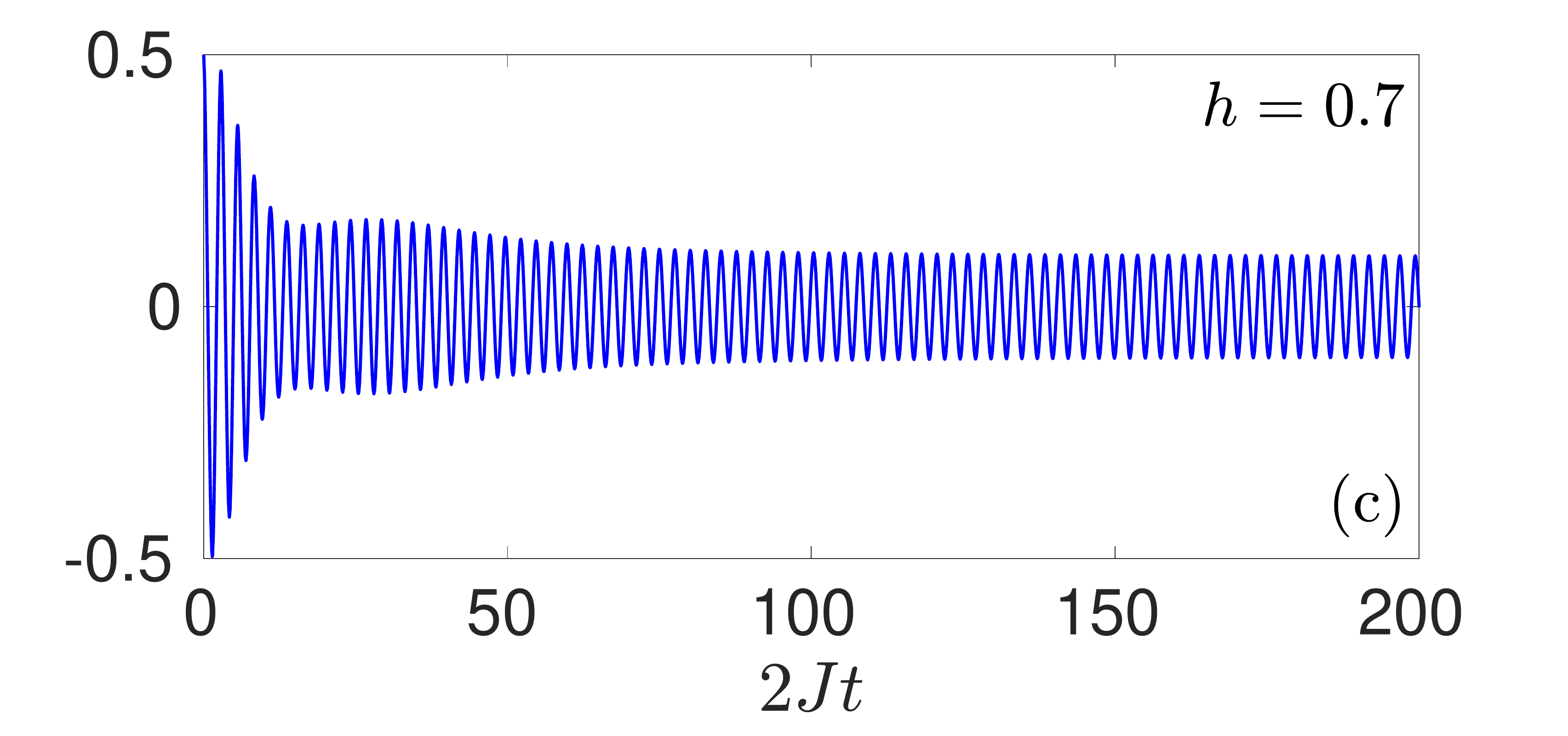}
    \end{minipage}
    \begin{minipage}{0.32\textwidth}
        \centering  
        \includegraphics[width=2.26in, height=1.8in]{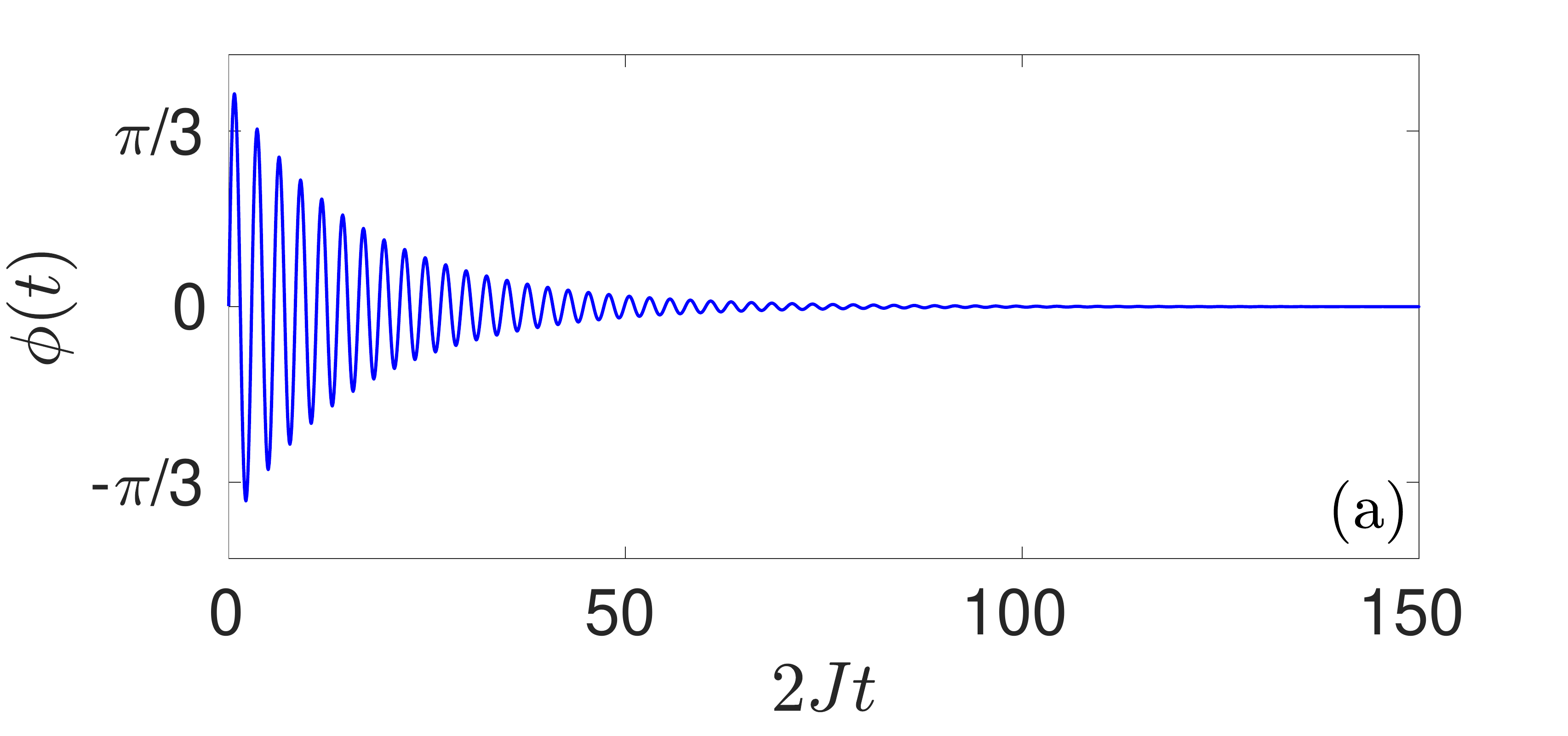}
    \end{minipage}
    \begin{minipage}{0.32\textwidth}
        \centering        
        \includegraphics[width=2.26in, height=1.8in]{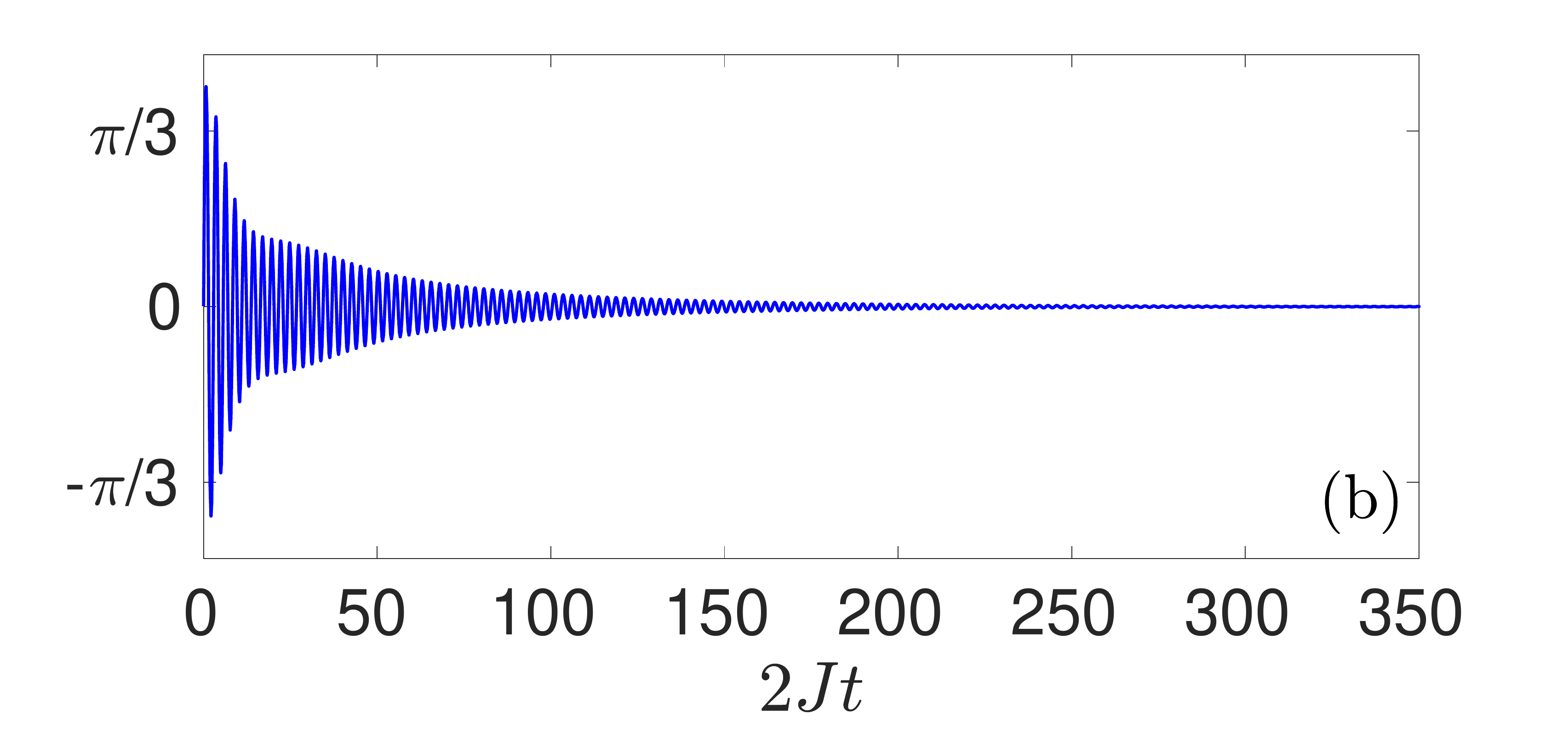}
    \end{minipage}
        \begin{minipage}{0.32\textwidth}
        \centering       
        \includegraphics[width=2.26in, height=1.8in]{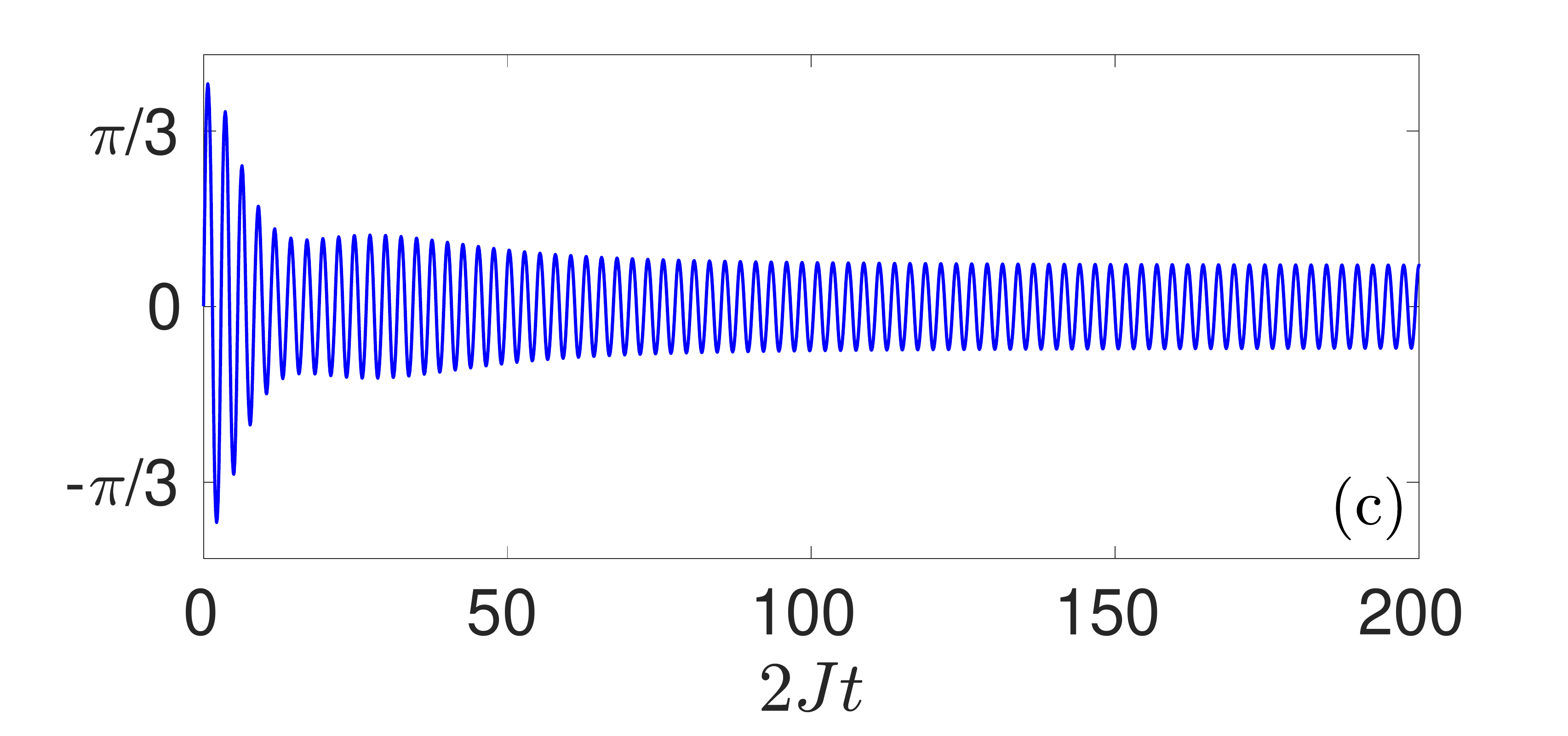}
    \end{minipage}
    \begin{minipage}{0.32\textwidth}
        \centering  
        \includegraphics[width=2.26in, height=1.8in]{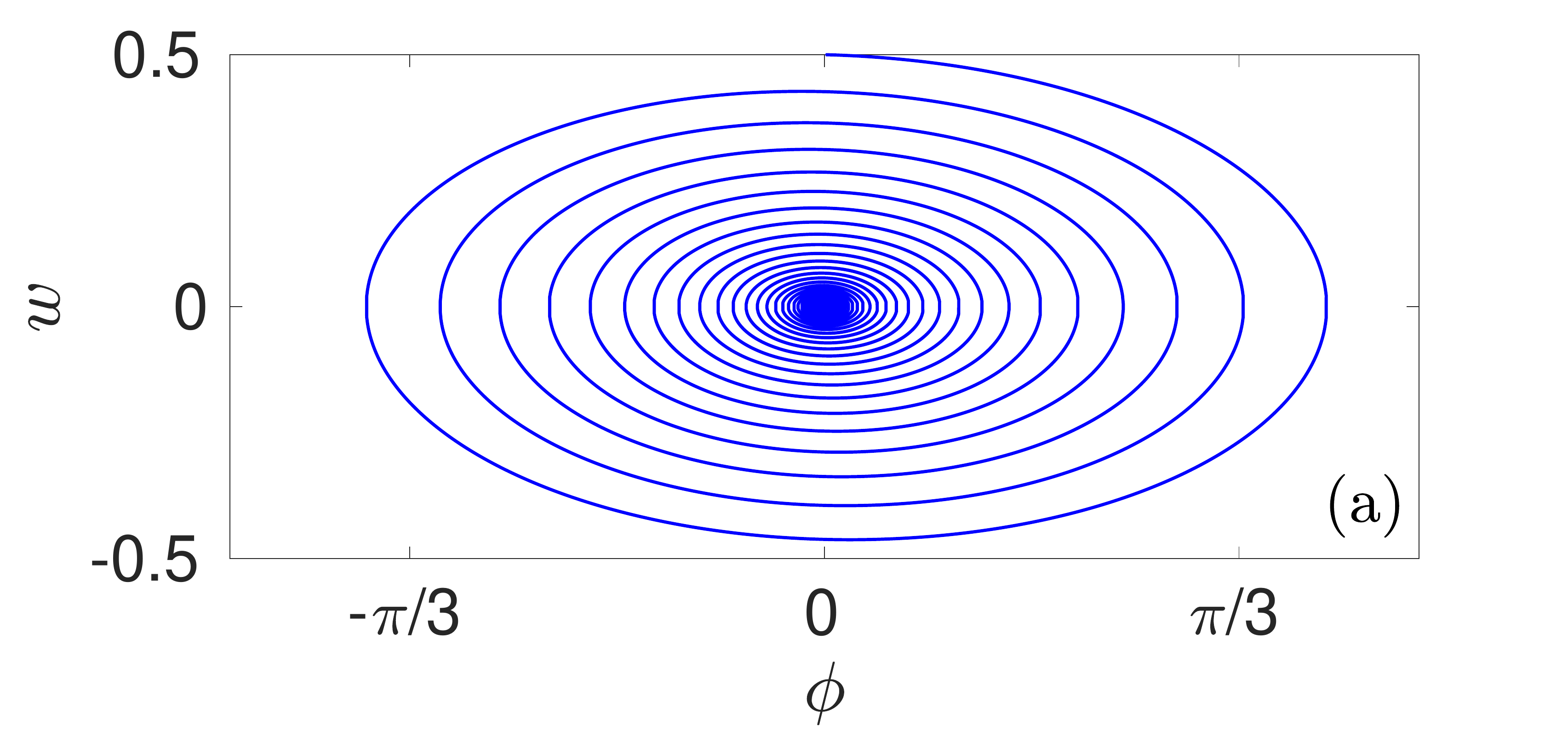}
    \end{minipage}
    \begin{minipage}{0.32\textwidth}
        \centering        
        \includegraphics[width=2.26in, height=1.8in]{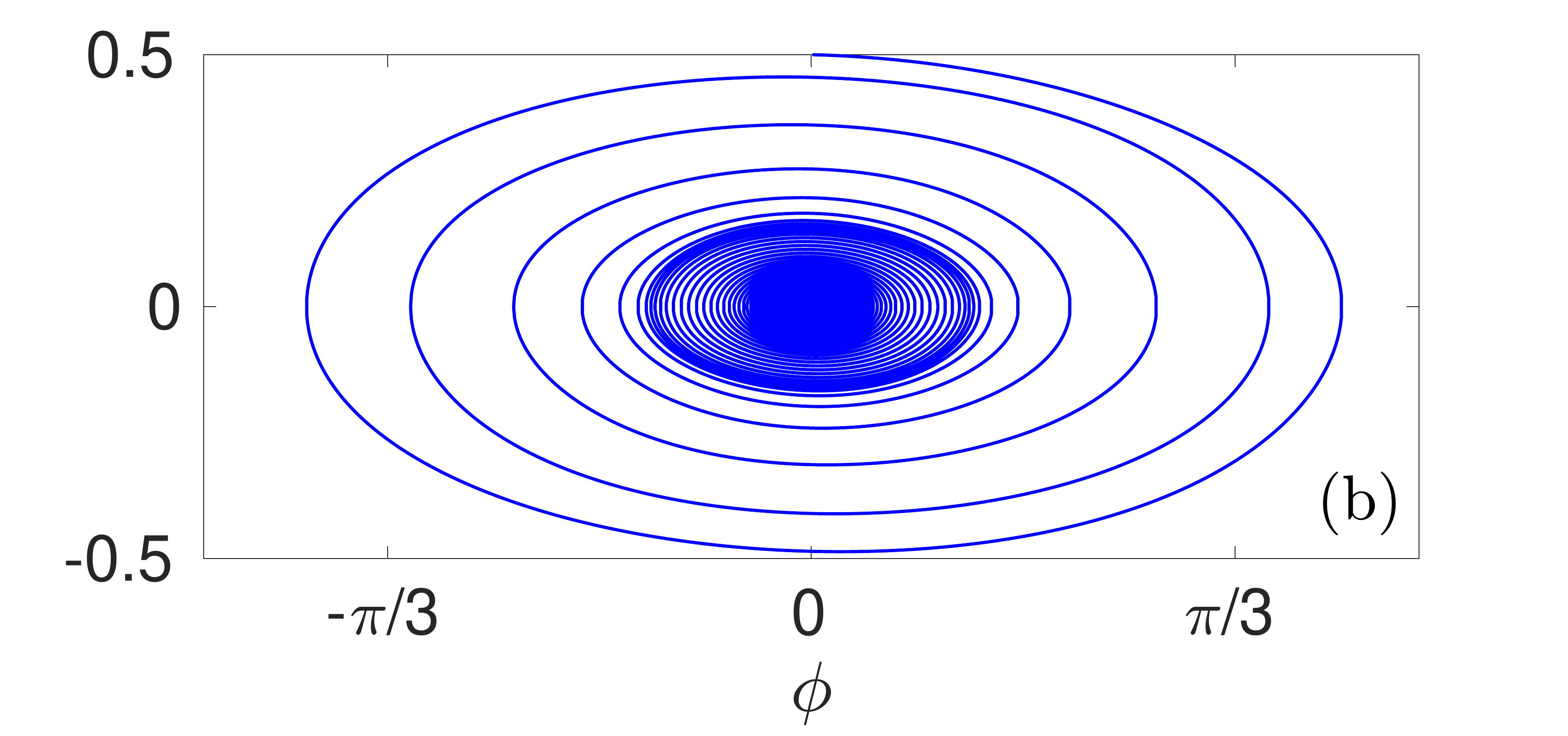}
    \end{minipage}
        \begin{minipage}{0.32\textwidth}
        \centering       
        \includegraphics[width=2.26in, height=1.8in]{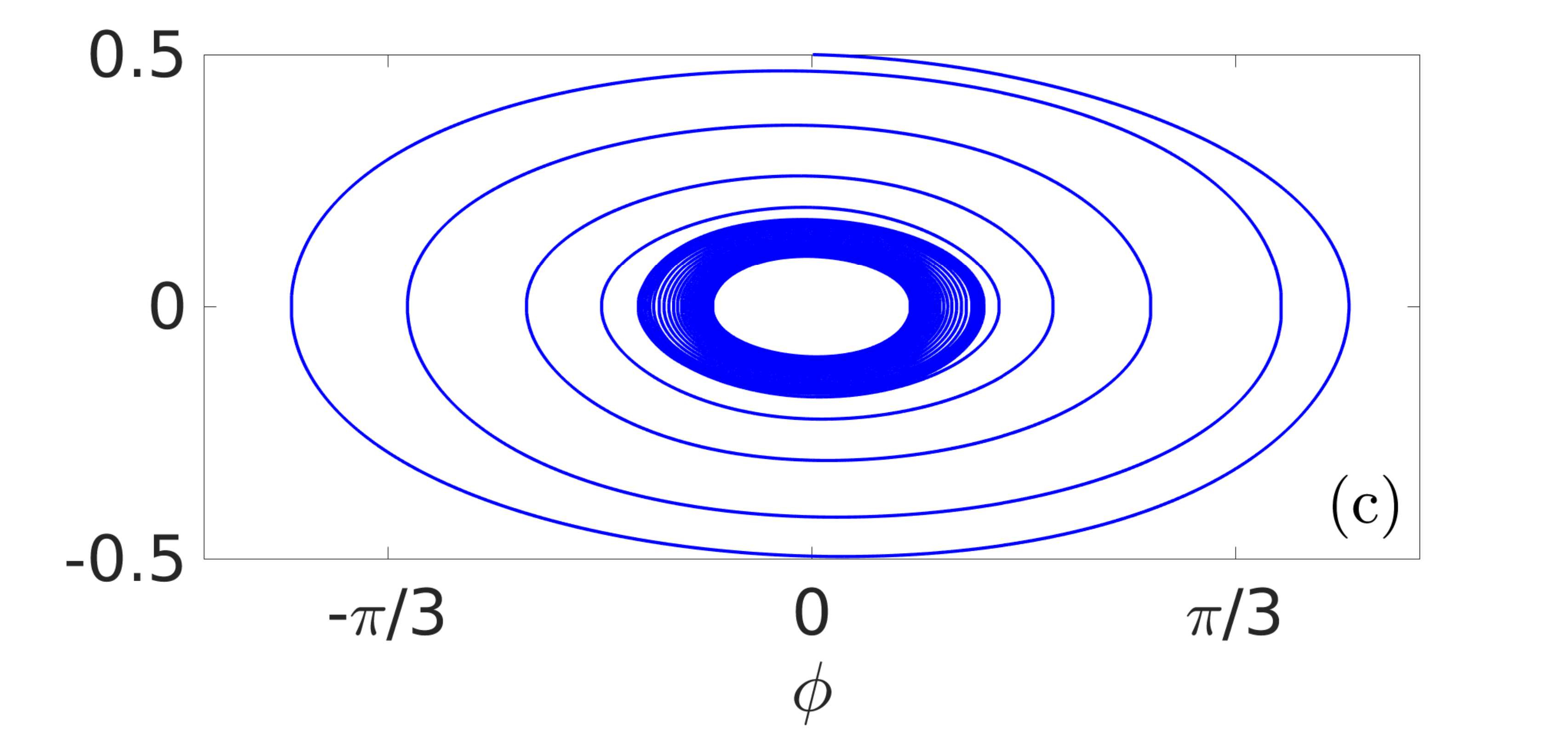}
    \end{minipage}
 \caption{\small Variation of the population imbalance $w(t)$ (first panel), phase difference $\phi(t)$ (second panel) as a function of dimensionless time $2Jt$ and phase-space trajectory (third panel) for different perturbation amplitudes (a) $h=0$ (b) $h=0.5$ and (c) $h=0.7$ with initial population imbalance $w(0)=0.5$, initial phase difference $\phi(0)=0$, $\frac{\eta}{N}=0.02$, $NU_{0}=0.24\hbar\omega_{z}$ with $N=2000$, $J=0.024\hbar\omega_{z}$, $\zeta=0.1$ and $\kappa=0.12$ in zero-phase mode.}
 \label{Fig1}
\end{figure}

\begin{figure}[h]
\centering
\includegraphics[width=3.5in, height=2.2in]{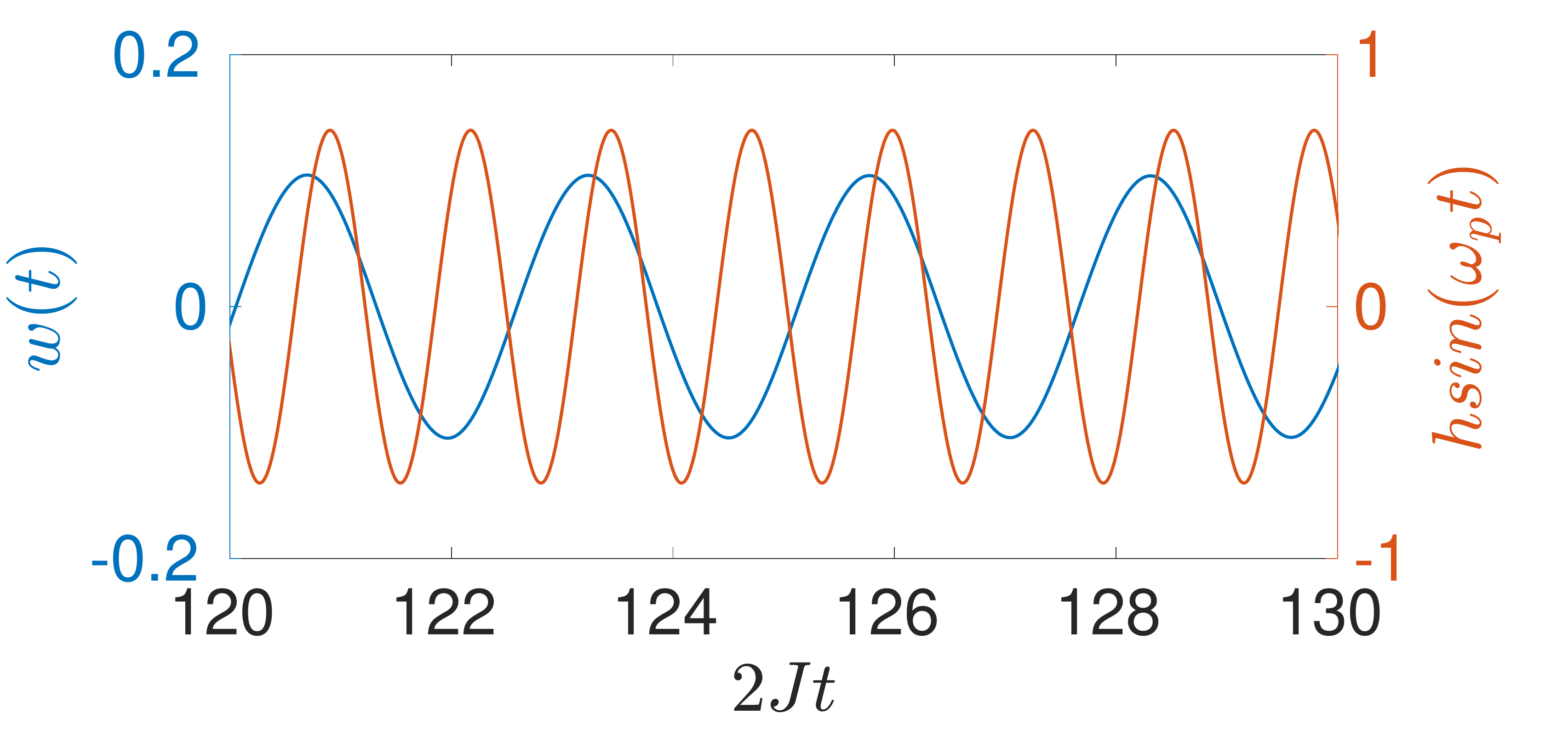}
\caption{\small Parametric oscillations in zero-phase mode where output signal $w(t)$ oscillates half of the input perturbation frequency $(\omega_{p})$ for parameter set as stated in Fig. \ref{Fig1}(c).}
\label{Fig2}
\end{figure}

To show the parametric oscillations numerically, we switch on the time-dependent perturbation $h\sin(\omega_{p}t)$ by setting $\omega_{p}=4.95$ and follow the oscillations for appropriate values of $h$ that lie just below and above of the threshold value for oscillation as determined by Eq. (16). Numerical simulation of Eqs. (3) and (4) under this condition shows that the system makes a transition from a steady state to a state of sustained oscillation. In the first panel of Fig. \ref{Fig1}, we show the variation of  $w(t)$ as a function of dimensionless time $2Jt$ for different perturbation amplitudes. It is observed that in the absence of the perturbation term $(h=0)$, the population imbalance approaches the stable steady state and the frequency of the oscillation governed by Eq. (18). Now, in presence of perturbation term $h=0.5$ $(h<h_{t})$, the system approaches to the stable steady state in the long time limit. When $h=0.7$, i.e, the threshold value $(h_{t}=0.6)$ of the perturbation term is crossed, the system exhibits the parametric oscillations around the stable steady state. Further with increase of the amplitude of the perturbation term, the dynamics remains the same. However this enhances the amplitude of the parametric oscillations. In the second panel of Fig. \ref{Fig1}, we plot the variation of the phase difference $\phi(t)$ as a function of dimensionless time. Below and above the threshold value, the dynamics of $\phi(t)$ shows similar behaviour as that of population imbalance. Finally, in the third panel of Fig. \ref{Fig1}, we show the phase-space trajectory with perturbation term being zero. The trajectory spirals towards the center with decreasing amplitude. On further increase of the value of the perturbation amplitude beyond the threshold, it moves over a circular path with finite radius, clearly depicting the parametric oscillation. In Fig. \ref{Fig2}, the wave profiles and the associated time periods of the time-dependent perturbation $hsin(\omega_{p}t)$ and the output response $w(t)$ are shown to demonstrate that $\omega_{p}$ matches well to $2\omega$ corresponding to the analytical estimate of the frequency obtained from Eq. (15).

\subsection{Mode transformation and MQST in running-phase mode}
\begin{figure}[h]
    \centering
    \begin{minipage}{.32\textwidth}
        \centering
        \includegraphics[width=2.26in, height=1.8in]{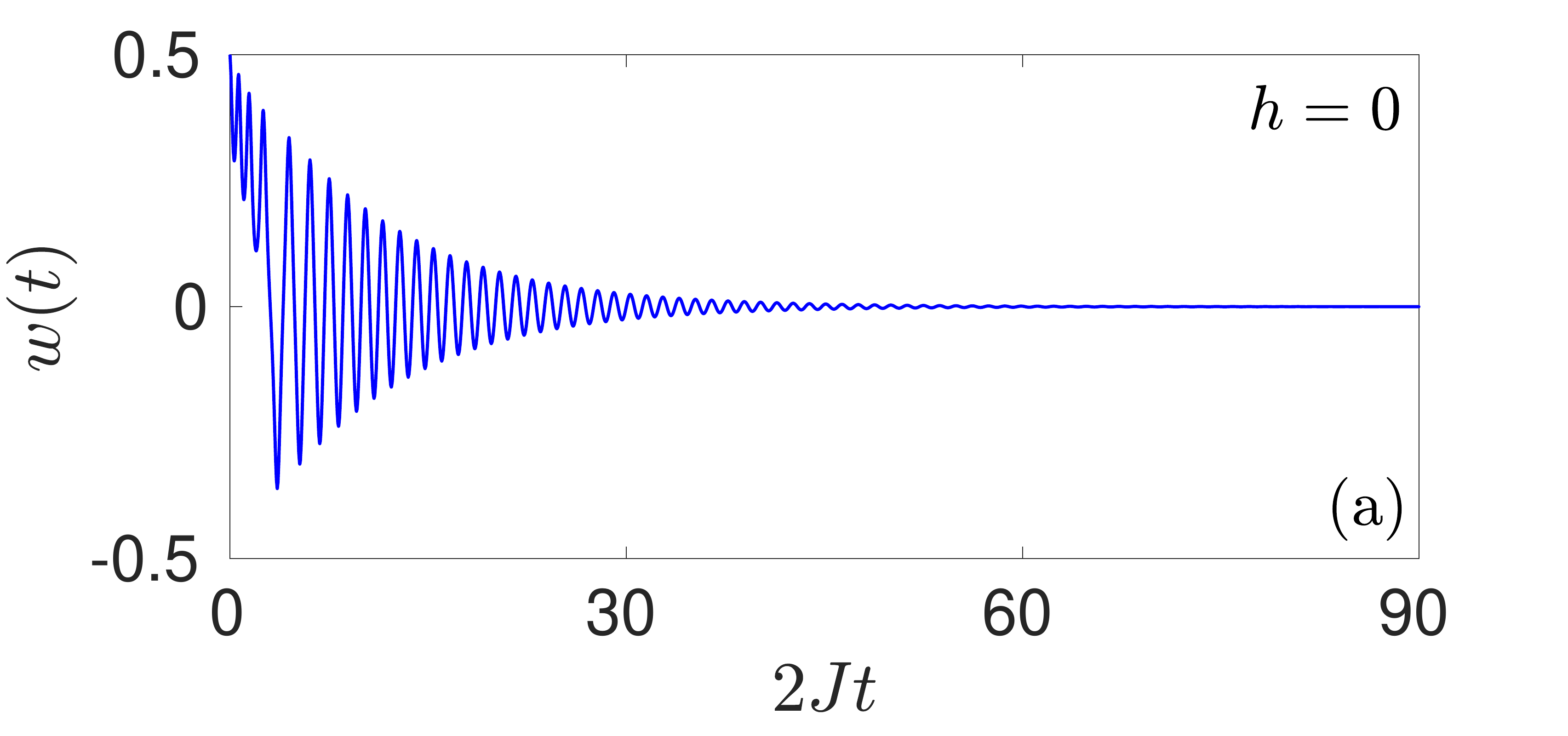}
    \end{minipage}
    \begin{minipage}{0.32\textwidth}
        \centering
        \includegraphics[width=2.26in, height=1.8in]{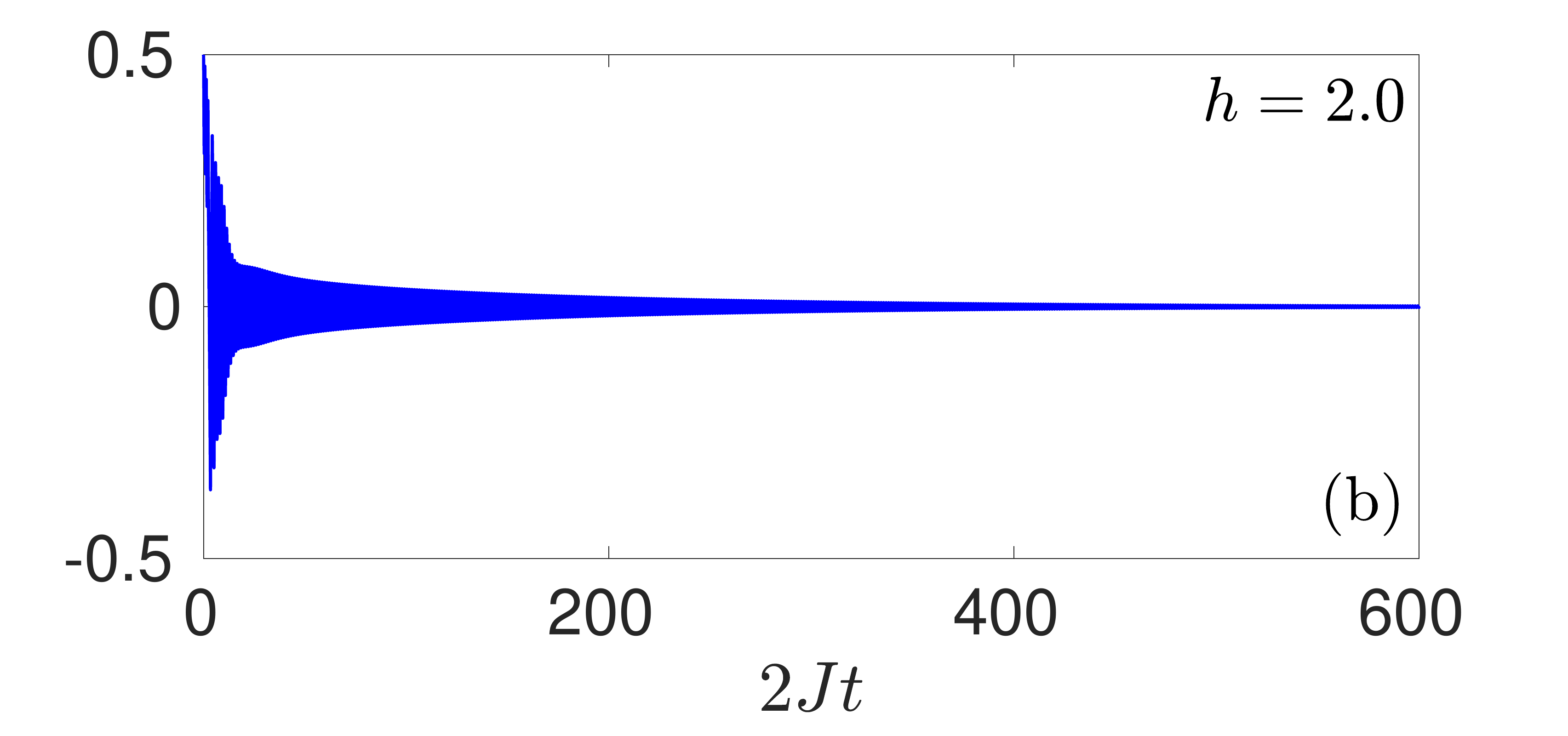}
    \end{minipage}
        \begin{minipage}{0.32\textwidth}
        \centering     
        \includegraphics[width=2.26in, height=1.8in]{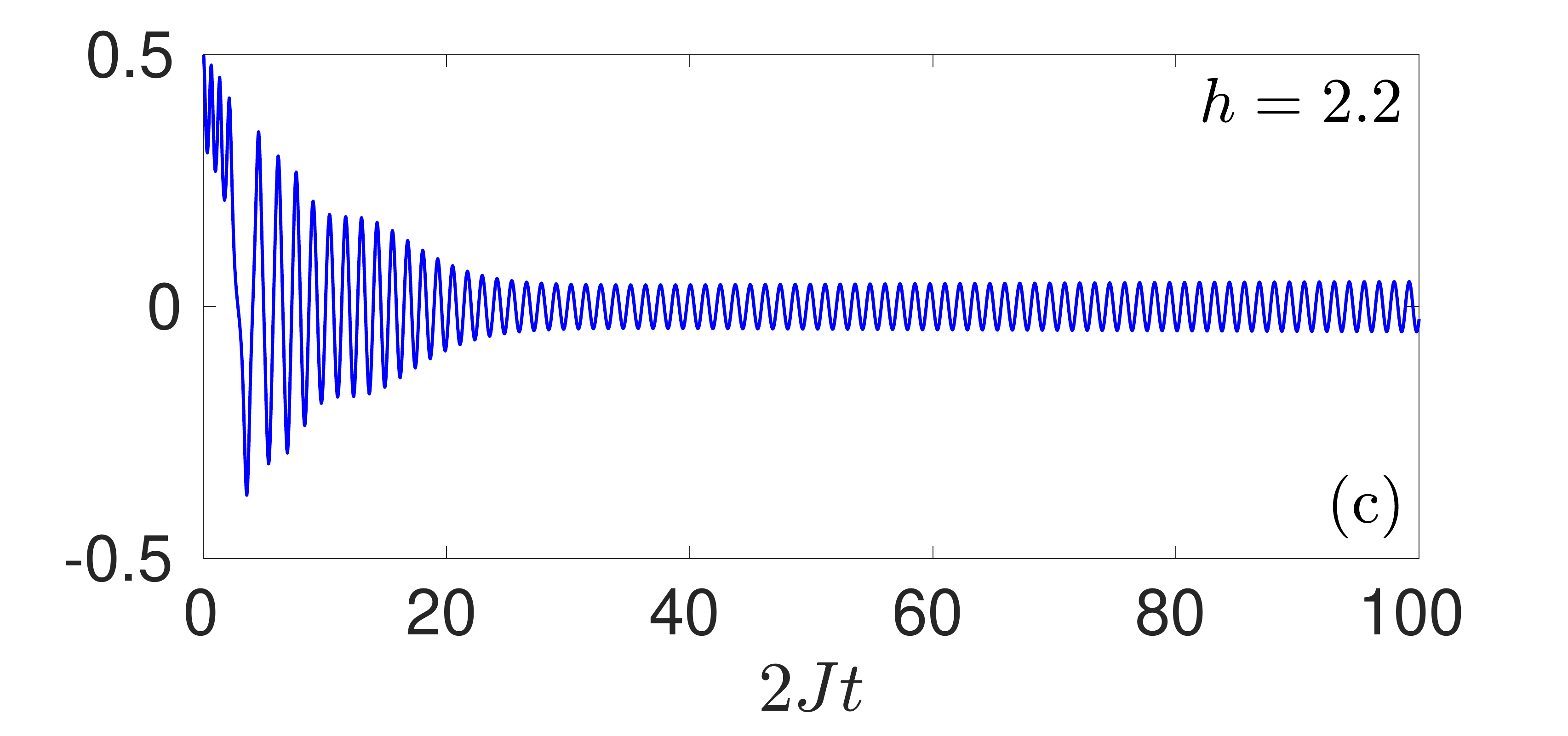}
    \end{minipage}
    \begin{minipage}{.32\textwidth}
        \centering  
        \includegraphics[width=2.26in, height=1.8in]{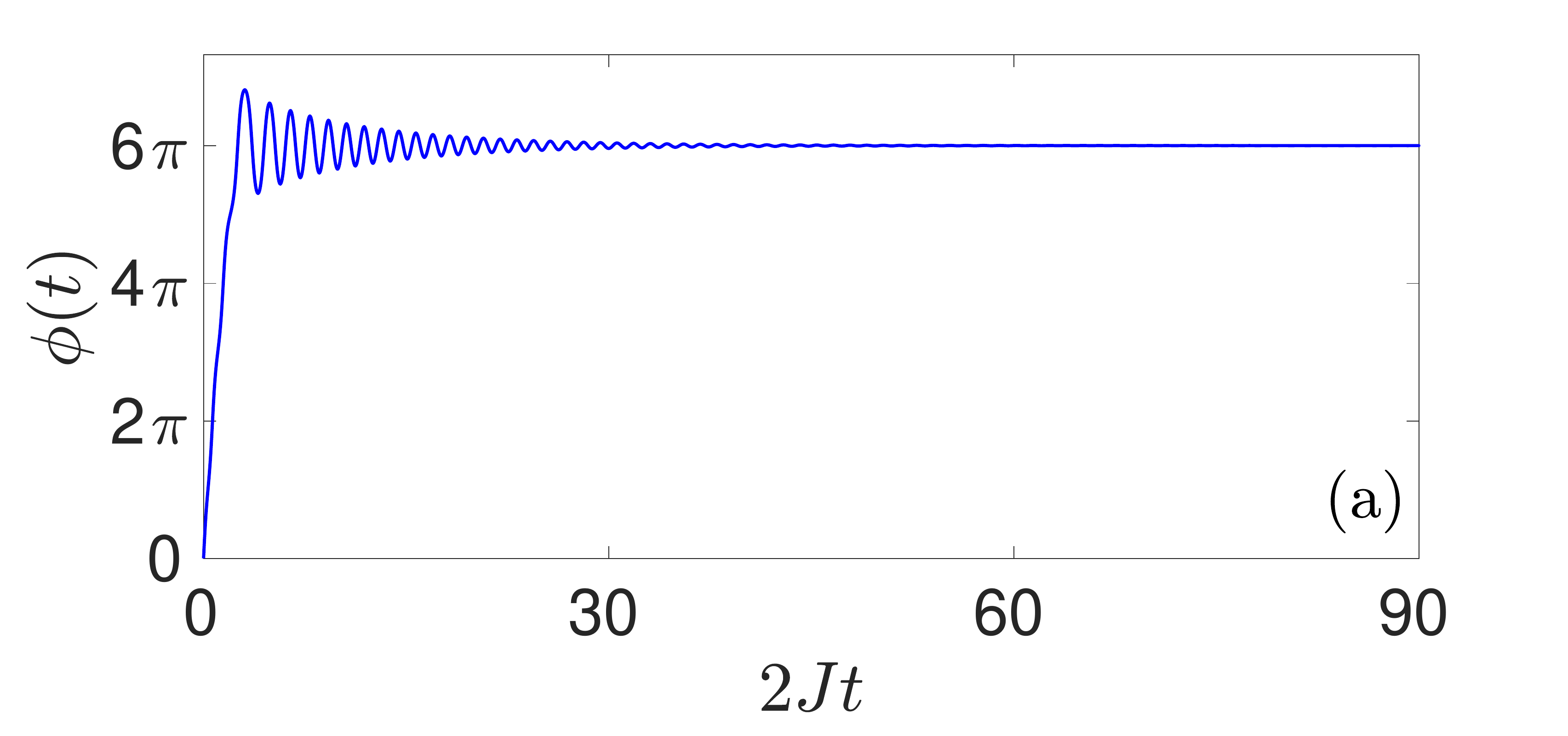}
    \end{minipage}
    \begin{minipage}{0.32\textwidth}
        \centering        
        \includegraphics[width=2.26in, height=1.8in]{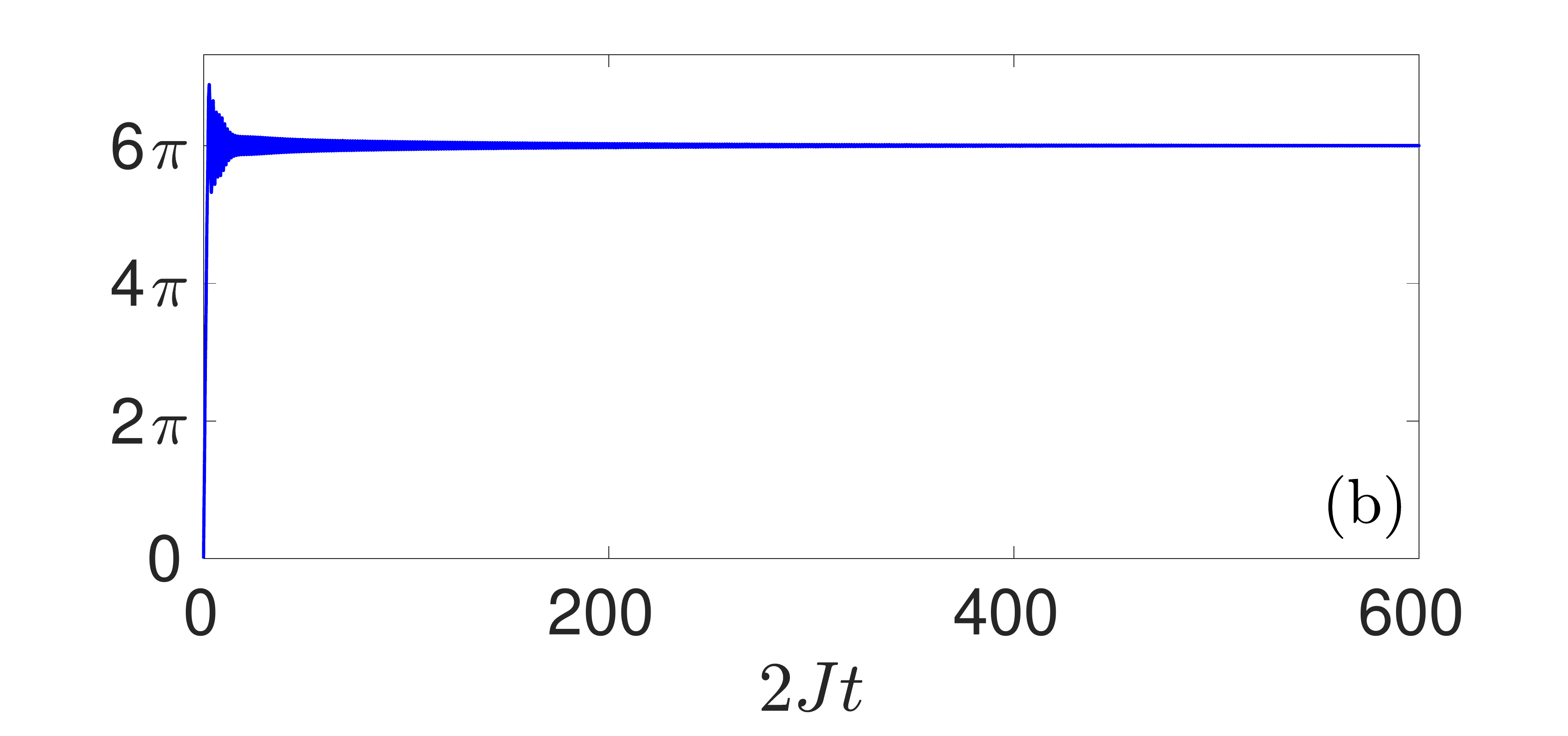}
    \end{minipage}
        \begin{minipage}{0.32\textwidth}
        \centering       
        \includegraphics[width=2.26in, height=1.8in]{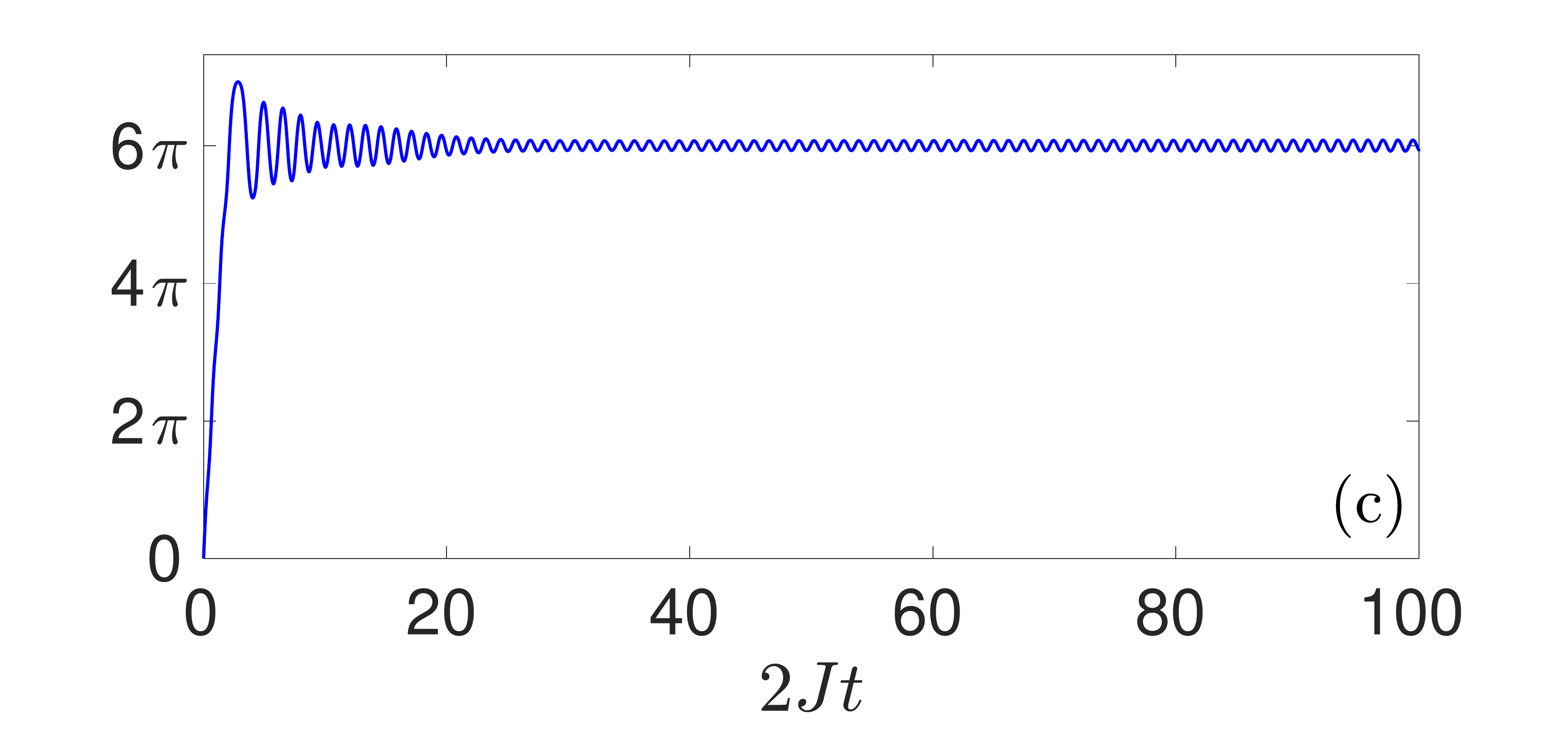}
    \end{minipage}
    \begin{minipage}{.32\textwidth}
        \centering  
        \includegraphics[width=2.26in, height=1.8in]{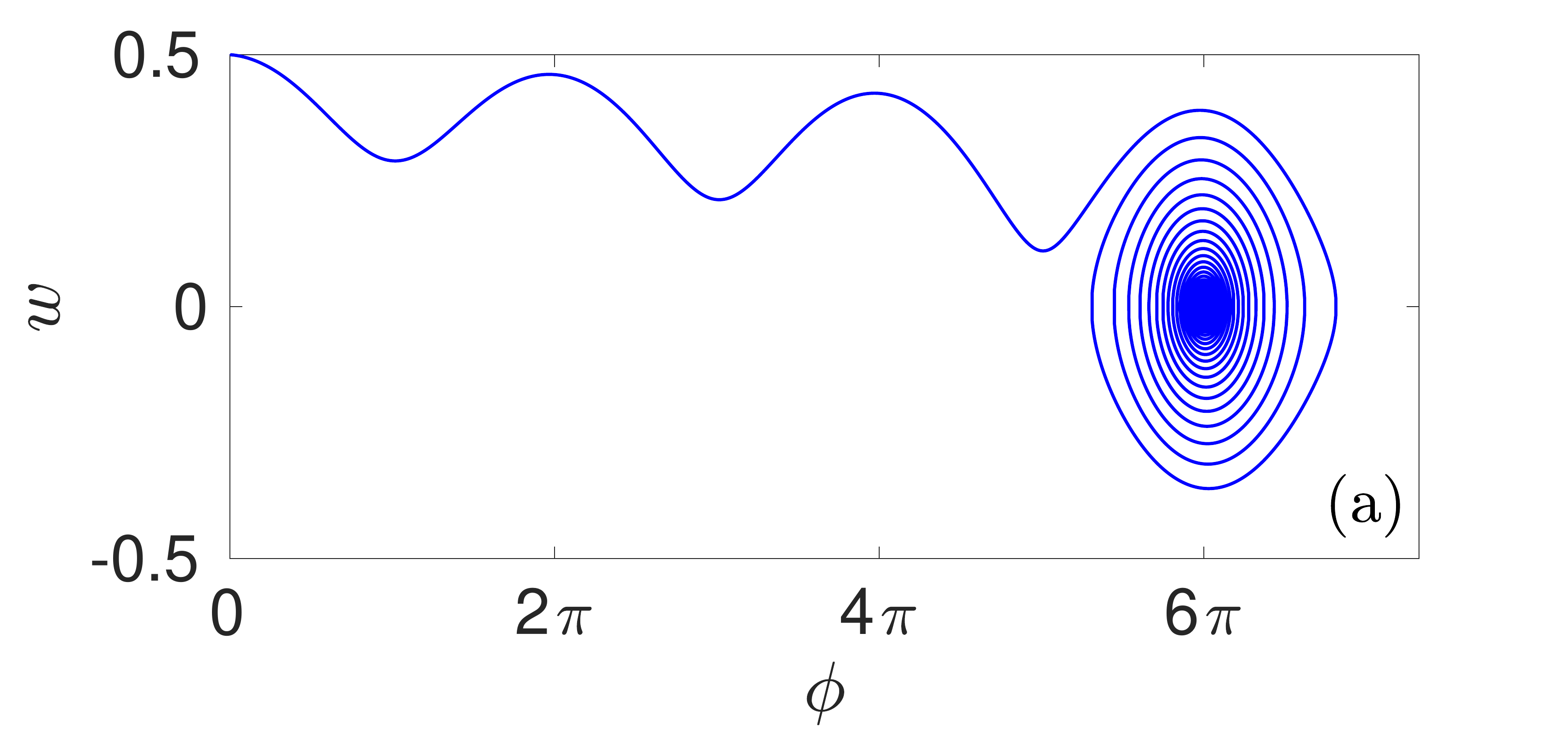}
    \end{minipage}
    \begin{minipage}{0.32\textwidth}
        \centering        
        \includegraphics[width=2.26in, height=1.8in]{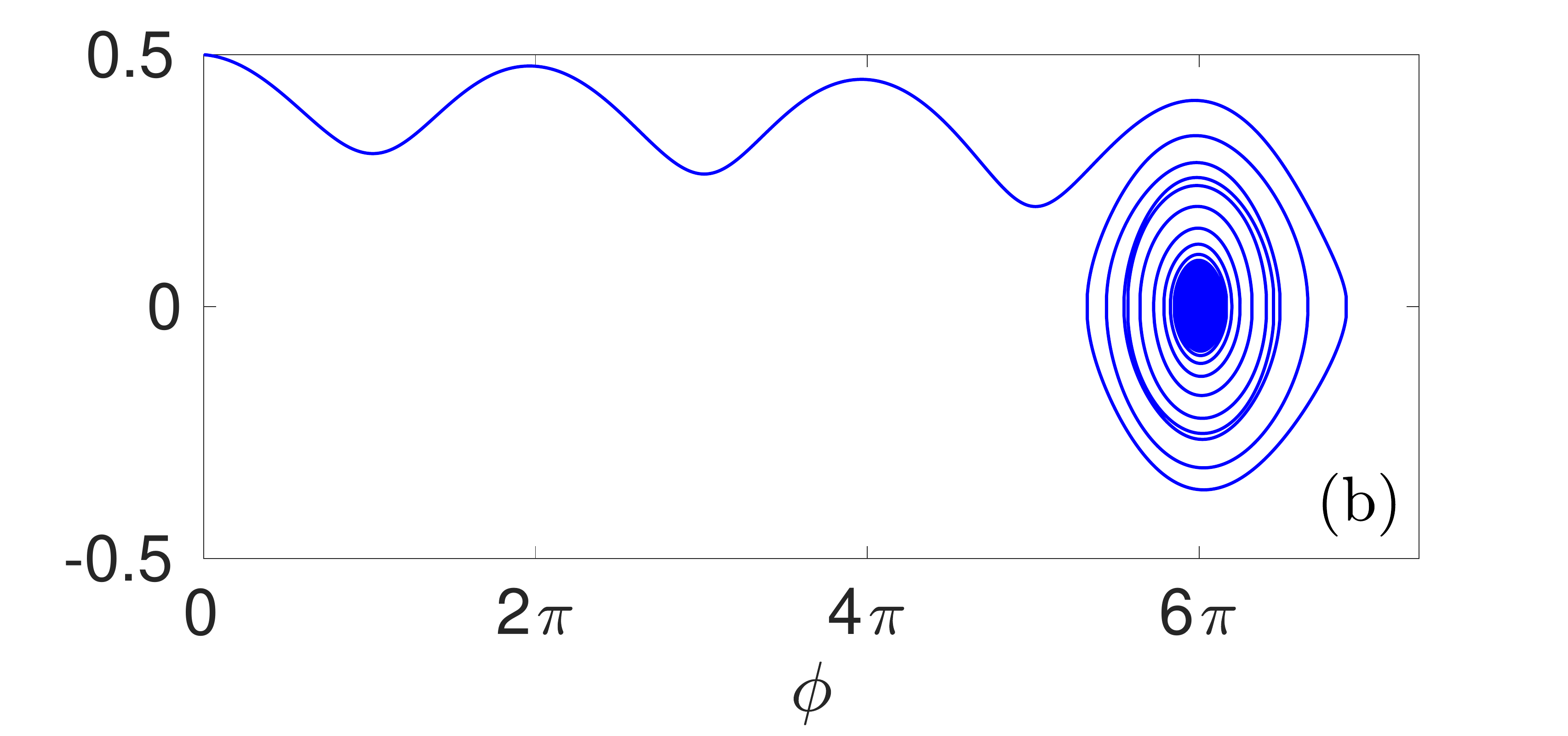}
    \end{minipage}
        \begin{minipage}{0.32\textwidth}
        \centering       
        \includegraphics[width=2.26in, height=1.8in]{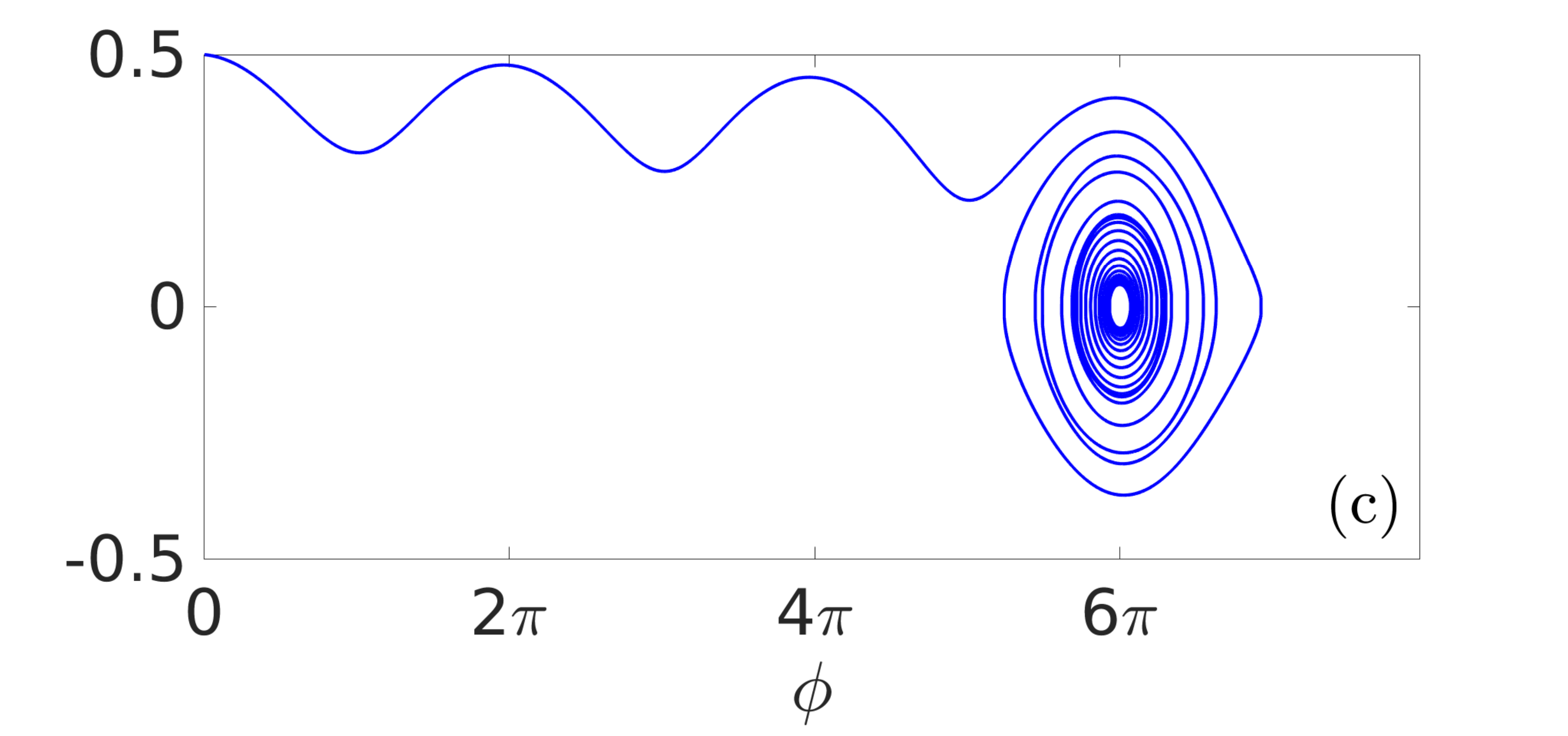}
    \end{minipage}
 \caption{\small Variation of $w(t)$ (first panel), $\phi(t)$ (second panel) as a function of $2Jt$ and phase-space trajectory (third panel) for (a) $h=0$ (b) $h=2.0$ and (c) $h=2.2$ with $w(0)=0.5$, $\phi(0)=0$, $\frac{\eta}{N}=0.008$, $\omega_{p}=10.20$, $NU_{0}=1.2\hbar\omega_{z}$ with $N=2000$, $J=0.024\hbar\omega_{z}$, $\zeta=0.08$ and $\kappa=0.20$  in running-phase mode.}
\label{Fig3}
\end{figure}

In the first panel of Fig. \ref{Fig3}, we show the variation of $w(t)$ as a function of dimensionless time $2Jt$ for different perturbation amplitudes. It is observed that in absence of the perturbation term $(h=0)$, the system decays to a self-trapping regime because of $\eta$ and then decreases to reach equilibrium. With increase of the value of perturbation amplitude $h=2.0$ $(h<h_{t})$, the dynamics shows similar decay profile in the long time. But when $h=2.2$, that is just above the $h_{t}=2.1$, at first the system decays to a self-trapping regime and then enters into a parametric oscillatory regime. Thus by increasing the perturbation amplitude above the threshold value one can realize a transition from self-trapping regime to parametric Josephson regime in dissipative BJJ. In the second panel of Fig. \ref{Fig3}, we show the time evolution of the phase difference $\phi(t)$ for different perturbation amplitudes. It is evident that when there is no perturbation term, $\phi(t)$ first increases with time and then enters to an oscillatory regime to settles down finally to $\phi=6\pi$, equivalent to the zero state that clearly describes the running-phase mode. Further, if we increase $h$ but for $h<h_{t}$, the system goes to the zero state after a long time. But when $h>h_{t}$, $\phi(t)$ originates from zero value, increases rapidly initially and then enters into the parametric oscillatory regime. We also plot the phase-space trajectory in the third panel of Fig. \ref{Fig3}. for the above perturbation amplitudes. It shows that the population imbalance decreases and spirals towards the $\phi=6\pi$ value with decrease in amplitude and beyond the threshold value, it oscillates to execute a circular motion with finite radius around $\phi=6\pi$ in a close loop.

\subsection{Phase slip and MQST in $\pi$-phase mode}
\begin{figure}[h]
    \centering
    \begin{minipage}{0.32\textwidth}
        \centering
        \includegraphics[width=2.24in, height=1.8in]{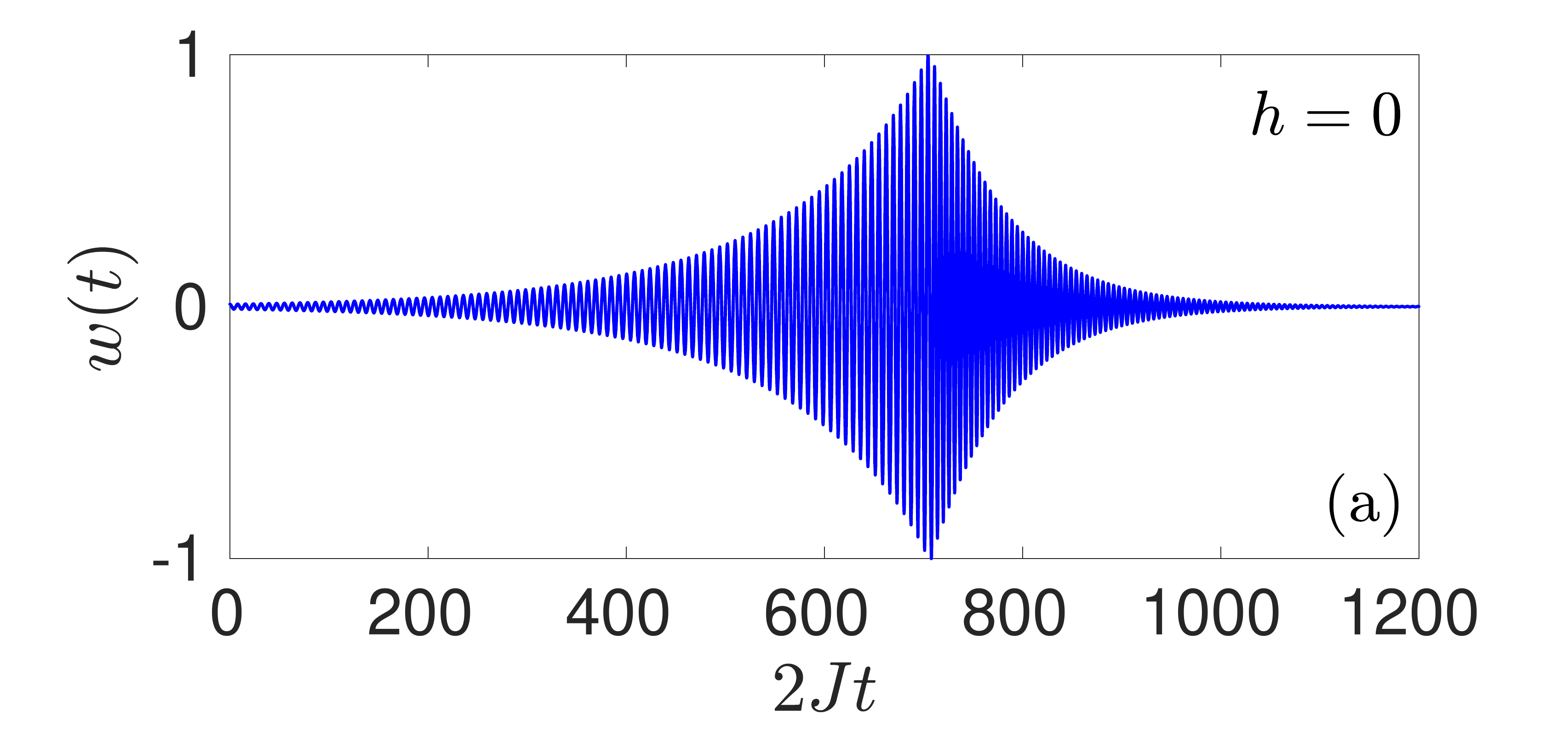}
    \end{minipage}
    \begin{minipage}{0.32\textwidth}
        \centering
        \includegraphics[width=2.24in, height=1.8in]{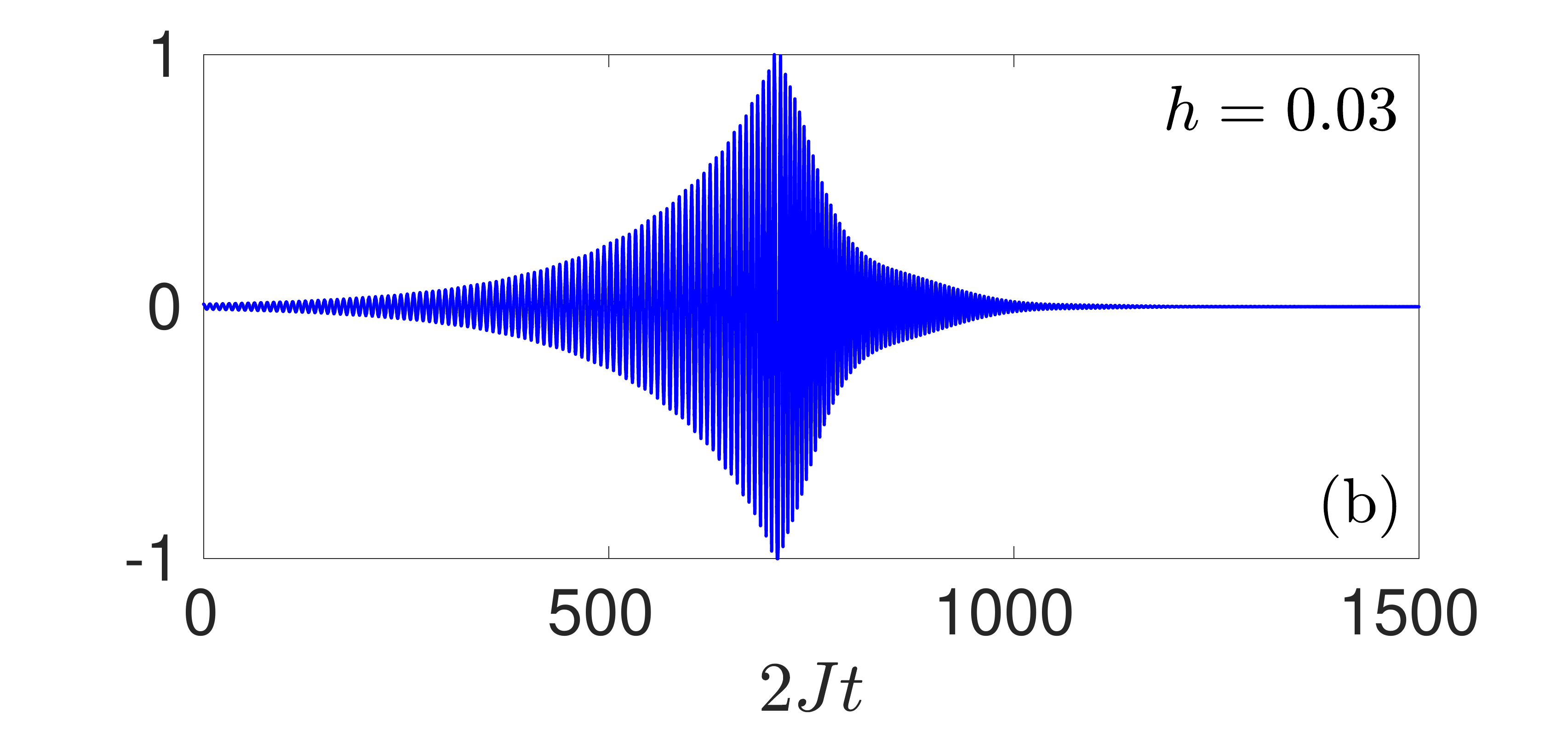}
    \end{minipage}
        \begin{minipage}{0.32\textwidth}
        \centering     
        \includegraphics[width=2.24in, height=1.8in]{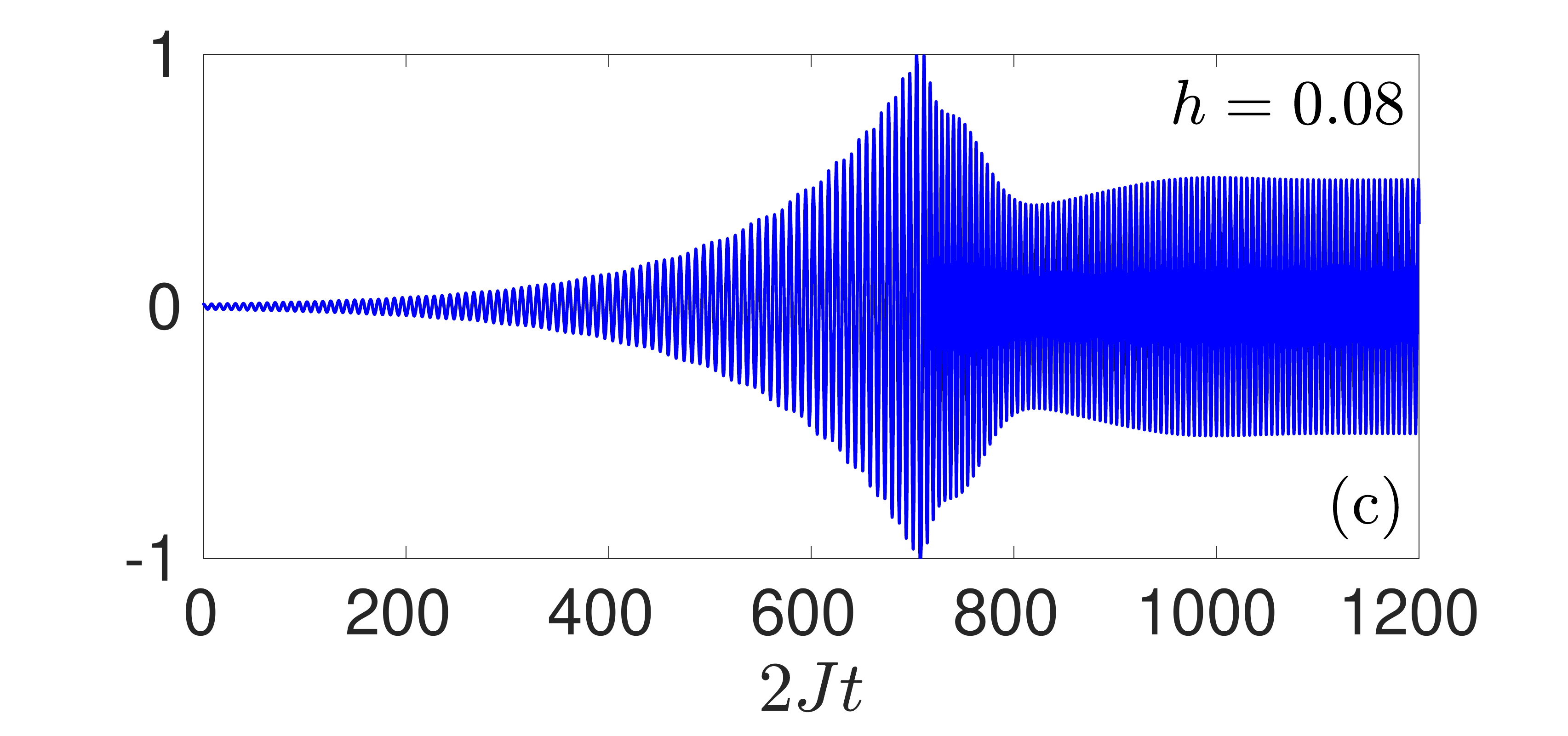}
    \end{minipage}
    \begin{minipage}{0.32\textwidth}
        \centering  
        \includegraphics[width=2.24in, height=1.8in]{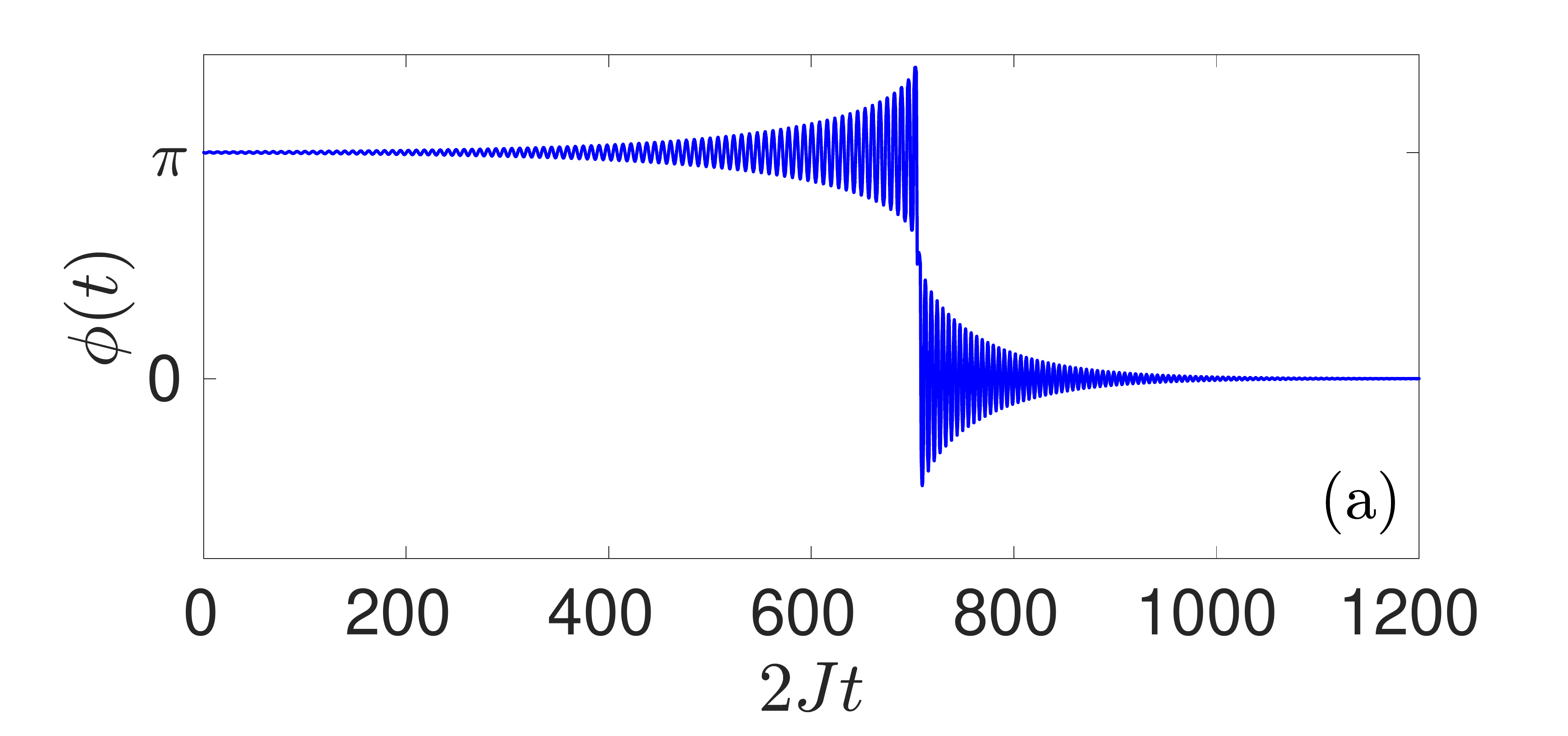}
    \end{minipage}
    \begin{minipage}{0.32\textwidth}
        \centering        
        \includegraphics[width=2.24in, height=1.8in]{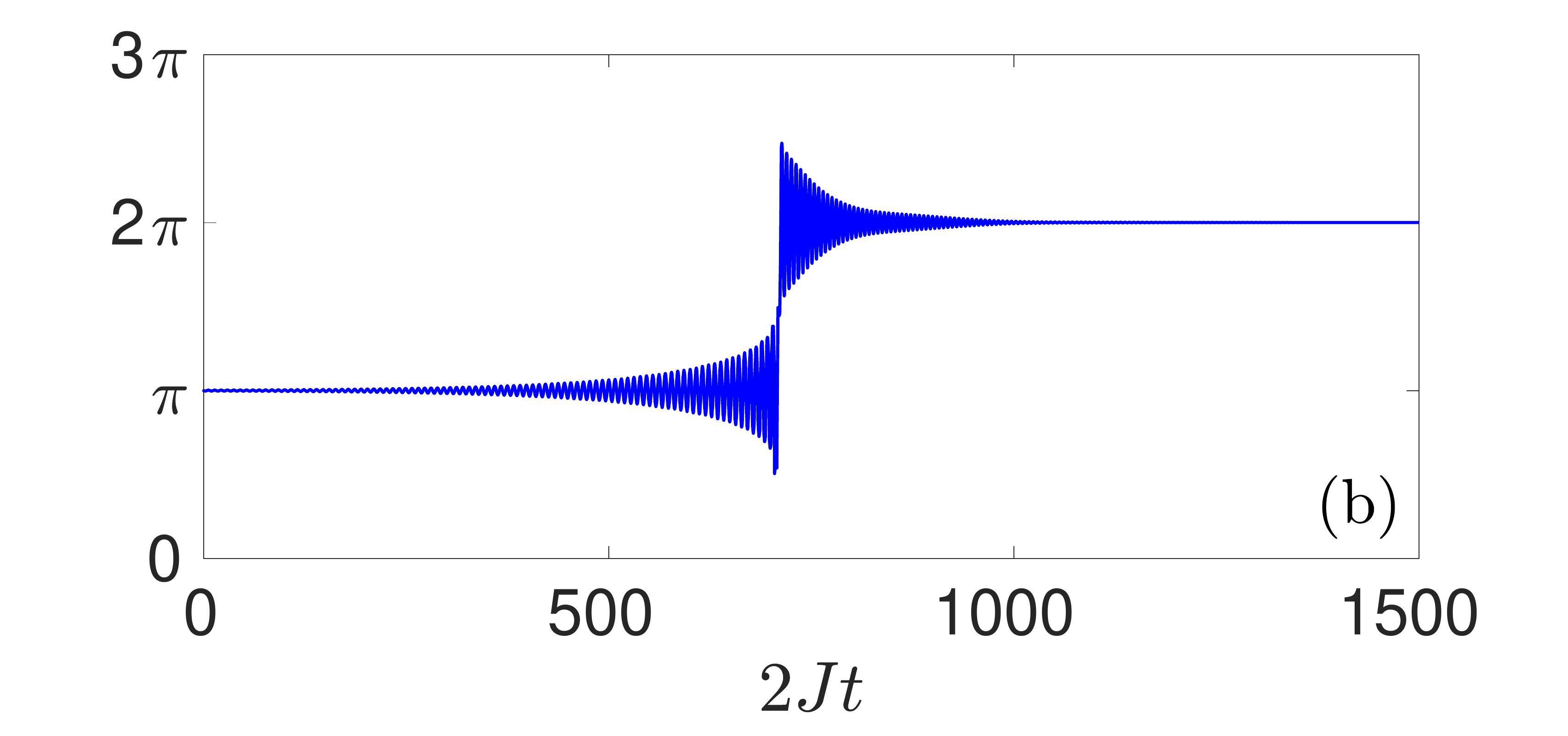}
    \end{minipage}
        \begin{minipage}{0.32\textwidth}
        \centering       
        \includegraphics[width=2.24in, height=1.8in]{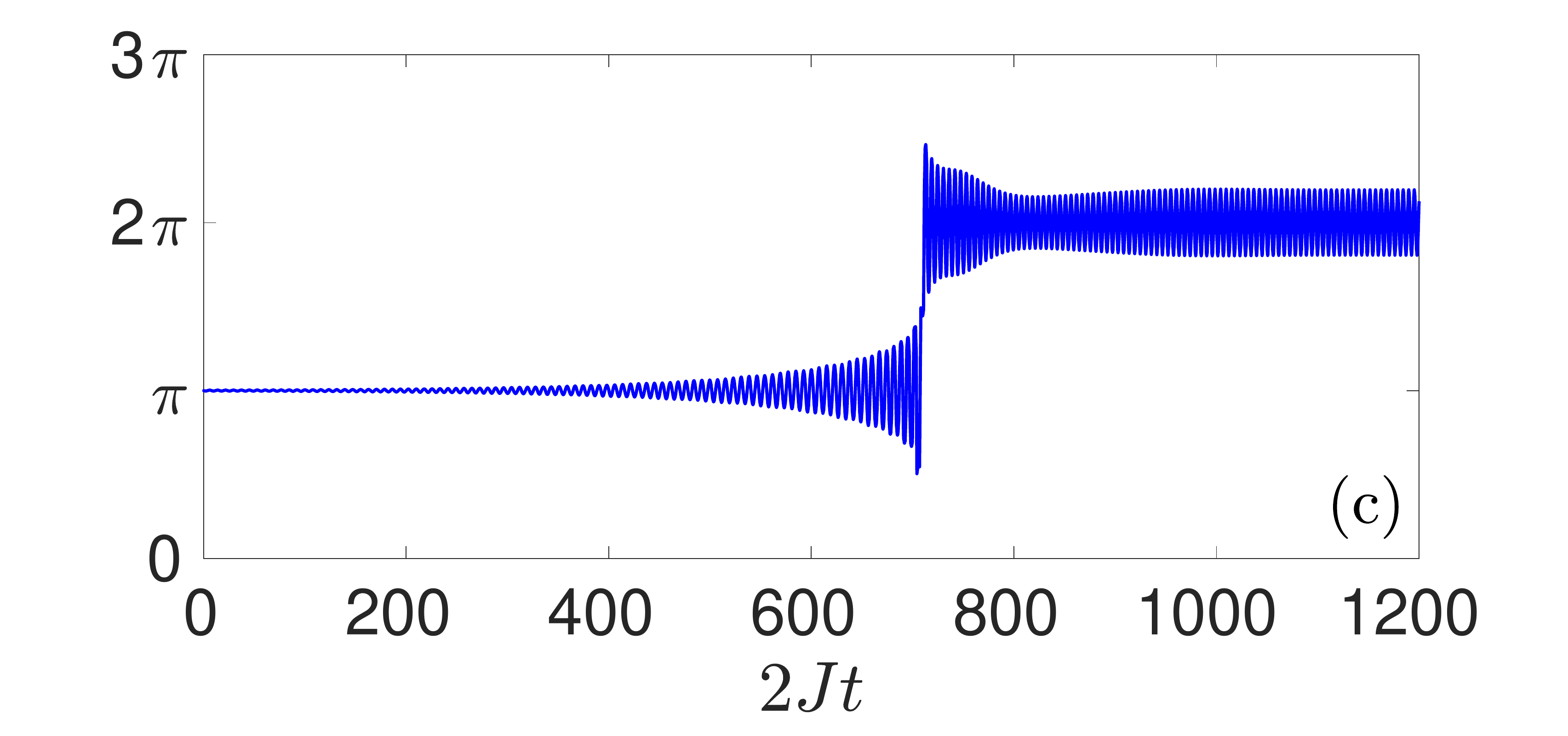}
    \end{minipage}
    \begin{minipage}{0.32\textwidth}
        \centering  
        \includegraphics[width=2.24in, height=1.8in]{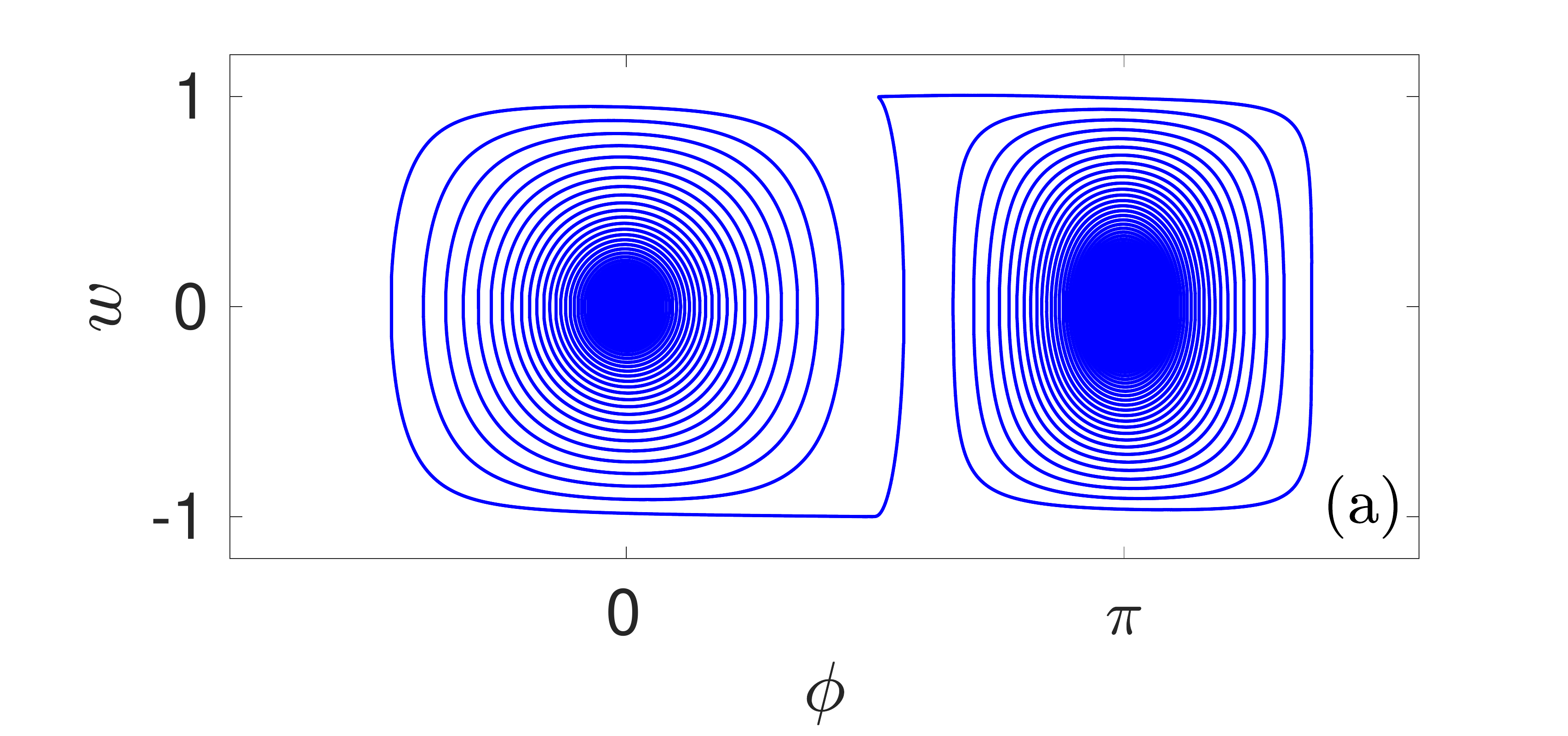}
    \end{minipage}
    \begin{minipage}{0.32\textwidth}
        \centering        
        \includegraphics[width=2.24in, height=1.8in]{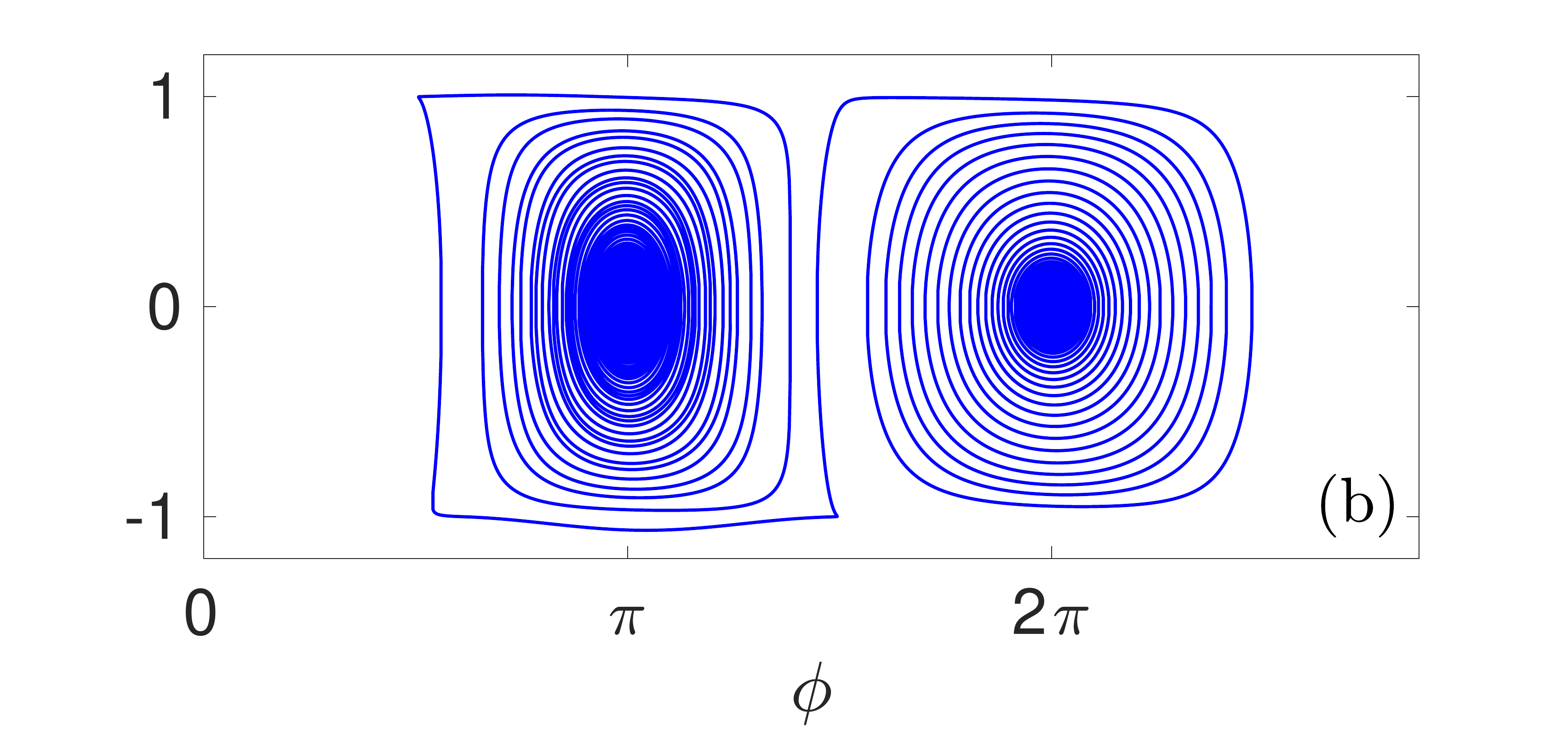}
    \end{minipage}
        \begin{minipage}{0.32\textwidth}
        \centering       
        \includegraphics[width=2.24in, height=1.8in]{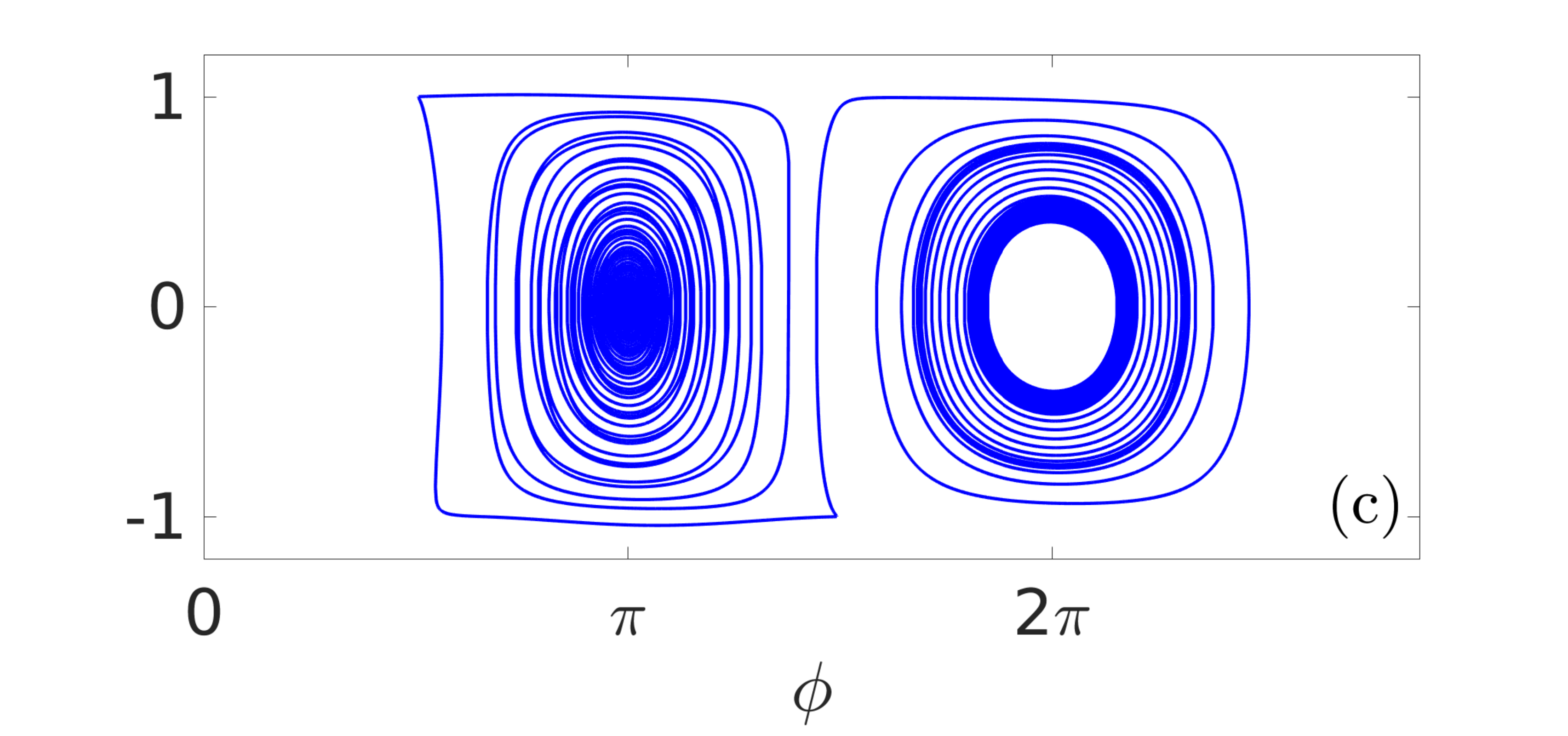}
    \end{minipage}
 \caption{\small Variation of $w(t)$ (first panel), $\phi(t)$ (second panel) as a function of $2Jt$ and phase-space trajectory (third panel) for (a) $h=0$ (b) $h=0.03$ and (c) $h=0.08$ with  $w(0)=0.01$, $\phi(0)=\pi$, $\frac{\eta}{N}=0.02$, $NU_{0}=0.017\hbar\omega_{z}$ with $N=2000$, $J=0.024\hbar\omega_{z}$, $\zeta=0.16$ and $\kappa=0.02$ in $\pi$-phase mode.}
 \label{Fig4}
\end{figure}

Numerical simulations of Eqs. (3) and (4) in $\pi$-phase mode shows that the amplitude of $w(t)$ increases with time until it saturates to unity. This implies that the underlying semi-classical approximation in Eqs. (3) and (4) breaks down and quantum fluctuations become important \cite{marino}. We, therefore follow the technique as described in Ref.  \cite{marino} to observe the behaviour of the system past the singularity. We also intend to enquire whether the parametric oscillation is possible or not in the $\pi$-phase mode. In the first panel of Fig. \ref{Fig4}, we plot the time evolution of the population imbalance for different perturbation amplitudes. At first, when there is no perturbation term, $w(t)$ oscillates with increasing amplitude over a time scale as stated in Eq. (22) and finally decays to reach equilibrium. This clearly describes the Josephson oscillations in dissipative BJJ and the characteristic frequency of this oscillation is governed by Eq. (21). Now, the time-dependent perturbation term is switched on by assigning a particular value of $\omega_{p}=2.30$. It is evident that when $h>h_{t}$ $(h_{t}=0.06)$, $w(t)$ oscillates with increasing amplitude before entering into the parametric oscillatory regime. To study the behaviour of the phase difference, we plot the time evolution of $\phi(t)$ for the above-mentioned perturbation amplitudes in the second panel of Fig. \ref{Fig4}. We see that when there is no perturbation term present, $\phi(t)$ originates from $\pi$ with increasing amplitude and after a certain time it jumps to the zero state with decrease in amplitude implying a phase slip with $\phi$ jumping by $\pi$. Further increase of the perturbation amplitude above the threshold value leads to oscillation with increasing amplitude around $\phi=\pi$ and after a certain time it jumps to the $\phi=2\pi$ state and then enters into the parametric oscillatory regime. In the third panel of Fig. \ref{Fig4}, we plot the phase-space trajectory for the above-mentioned perturbation terms. In the absence of perturbation term, it shows that the population imbalance spirals outwards around $\phi=\pi$ and after the phase slip occurs around $\phi=\frac{\pi}{2}$, it spirals towards the center with decrease in amplitude around $\phi=0$. When the threshold value is crossed $w$ spirals outward around $\phi=\pi$ and after phase slip around $\phi=\frac{3\pi}{2}$, it oscillates over a closed phase space loop with finite radius around $\phi=2\pi$. 

\begin{figure}[h]
    \centering
    \begin{minipage}{.32\textwidth}
        \centering
        \includegraphics[width=2.26in, height=1.8in]{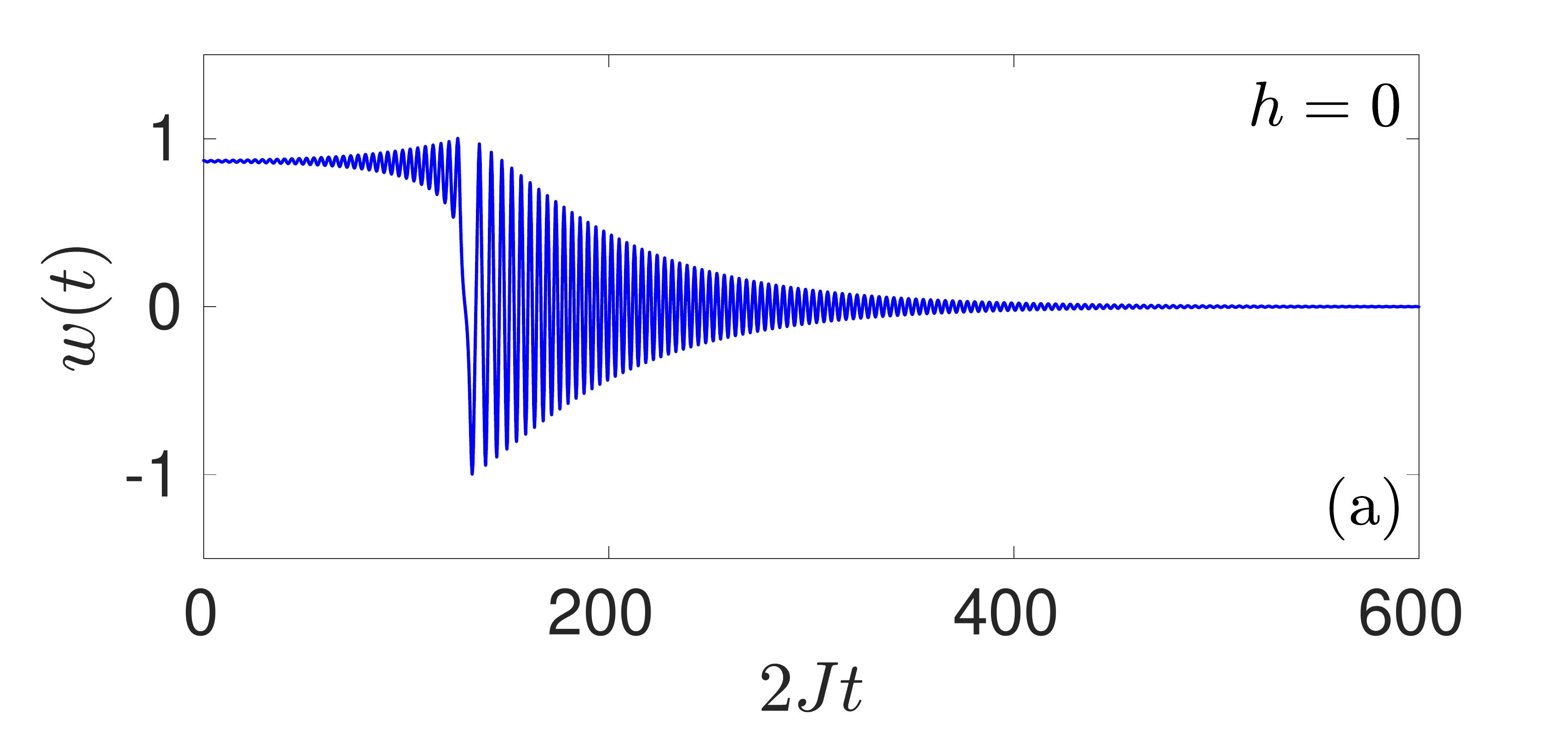}
    \end{minipage}
    \begin{minipage}{0.32\textwidth}
        \centering
        \includegraphics[width=2.26in, height=1.8in]{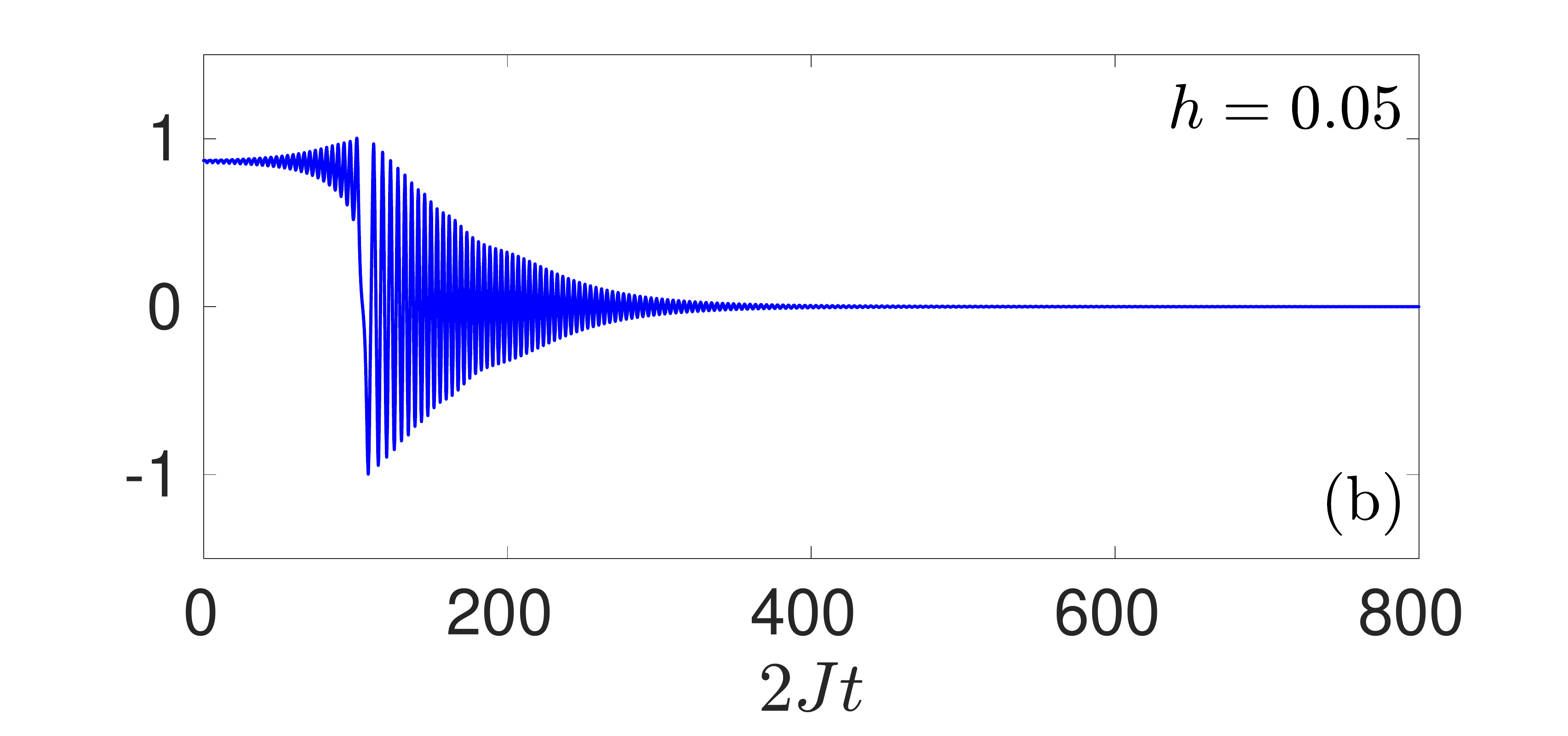}
    \end{minipage}
        \begin{minipage}{0.32\textwidth}
        \centering     
        \includegraphics[width=2.26in, height=1.8in]{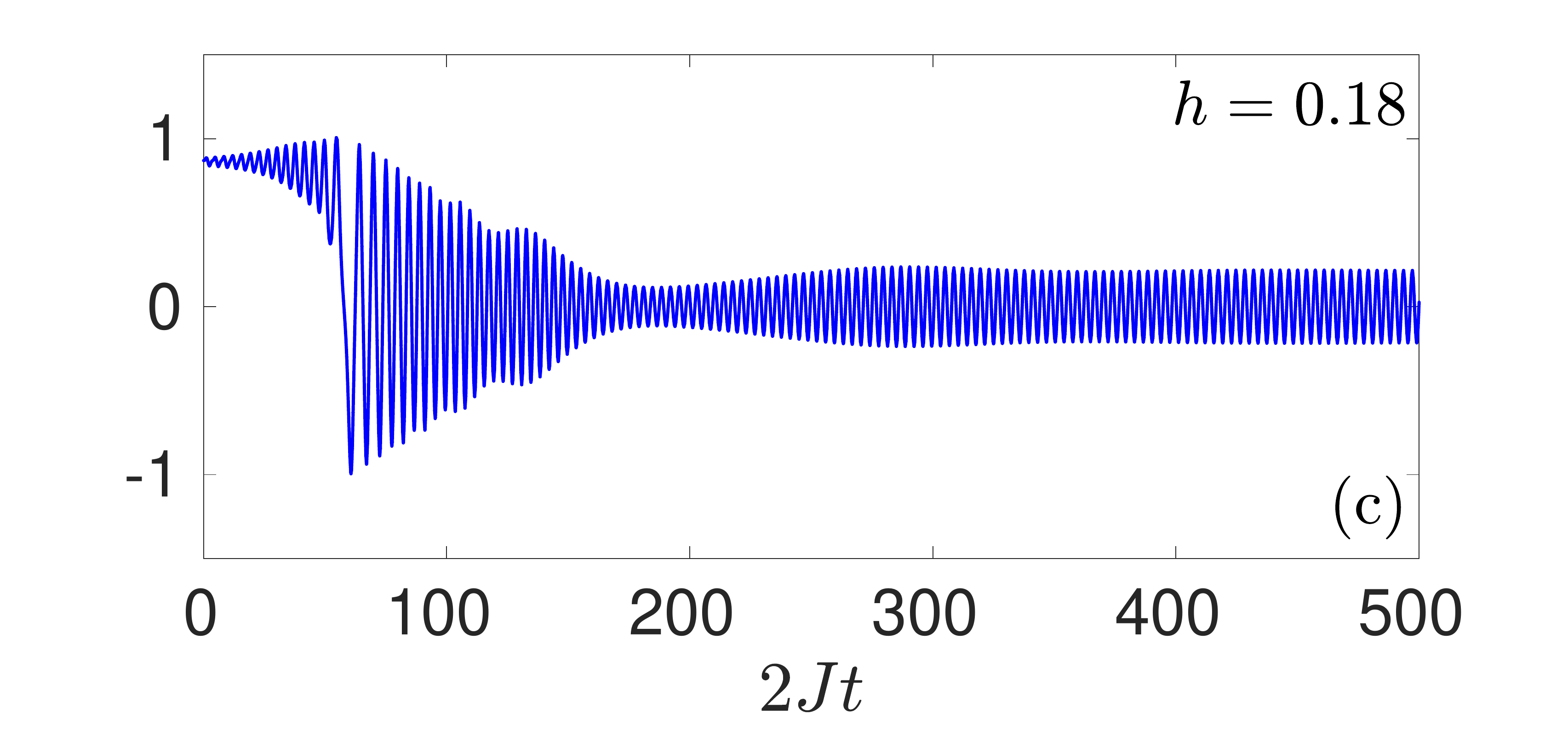}
    \end{minipage}
    \begin{minipage}{.32\textwidth}
        \centering  
        \includegraphics[width=2.26in, height=1.8in]{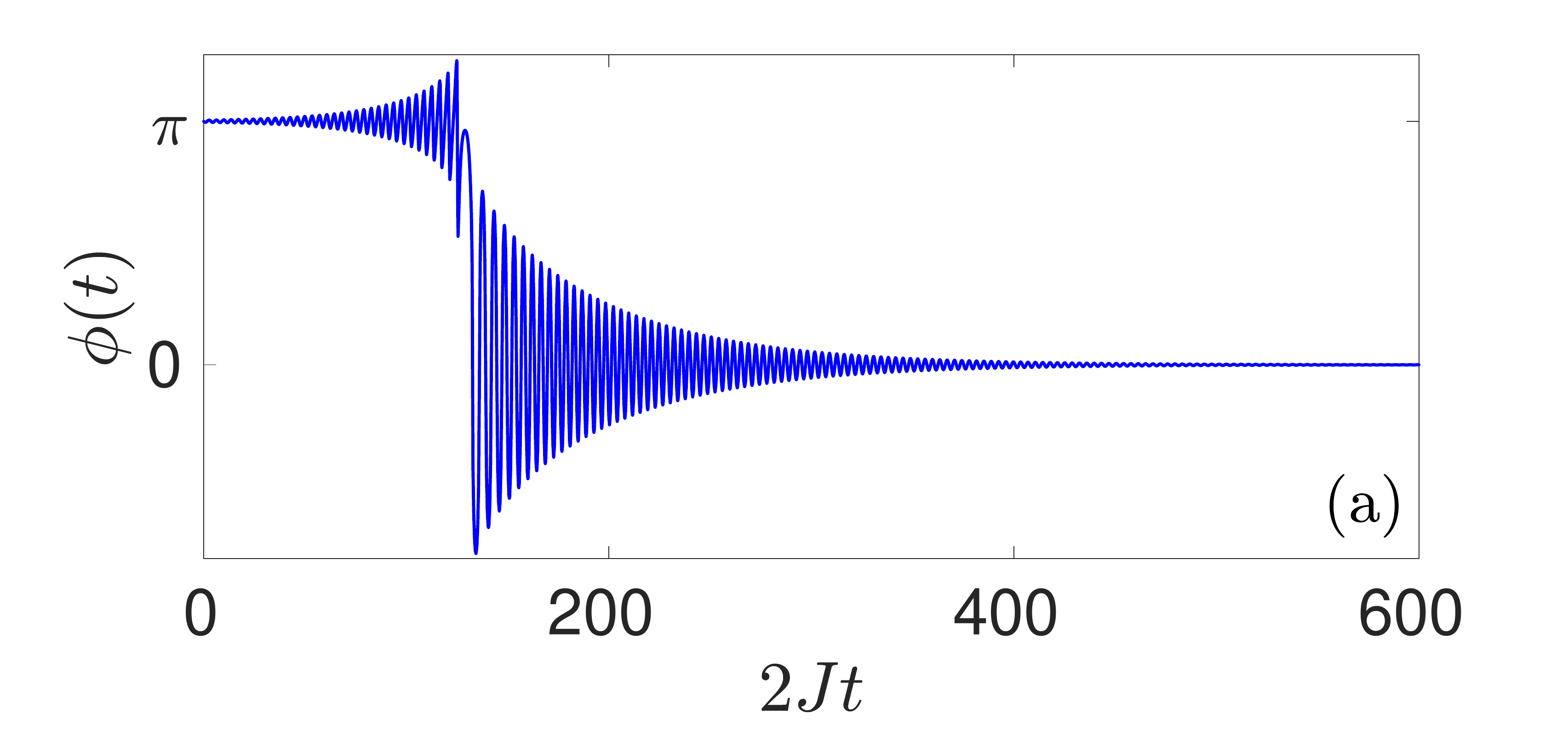}
    \end{minipage}
    \begin{minipage}{0.32\textwidth}
        \centering        
        \includegraphics[width=2.26in, height=1.8in]{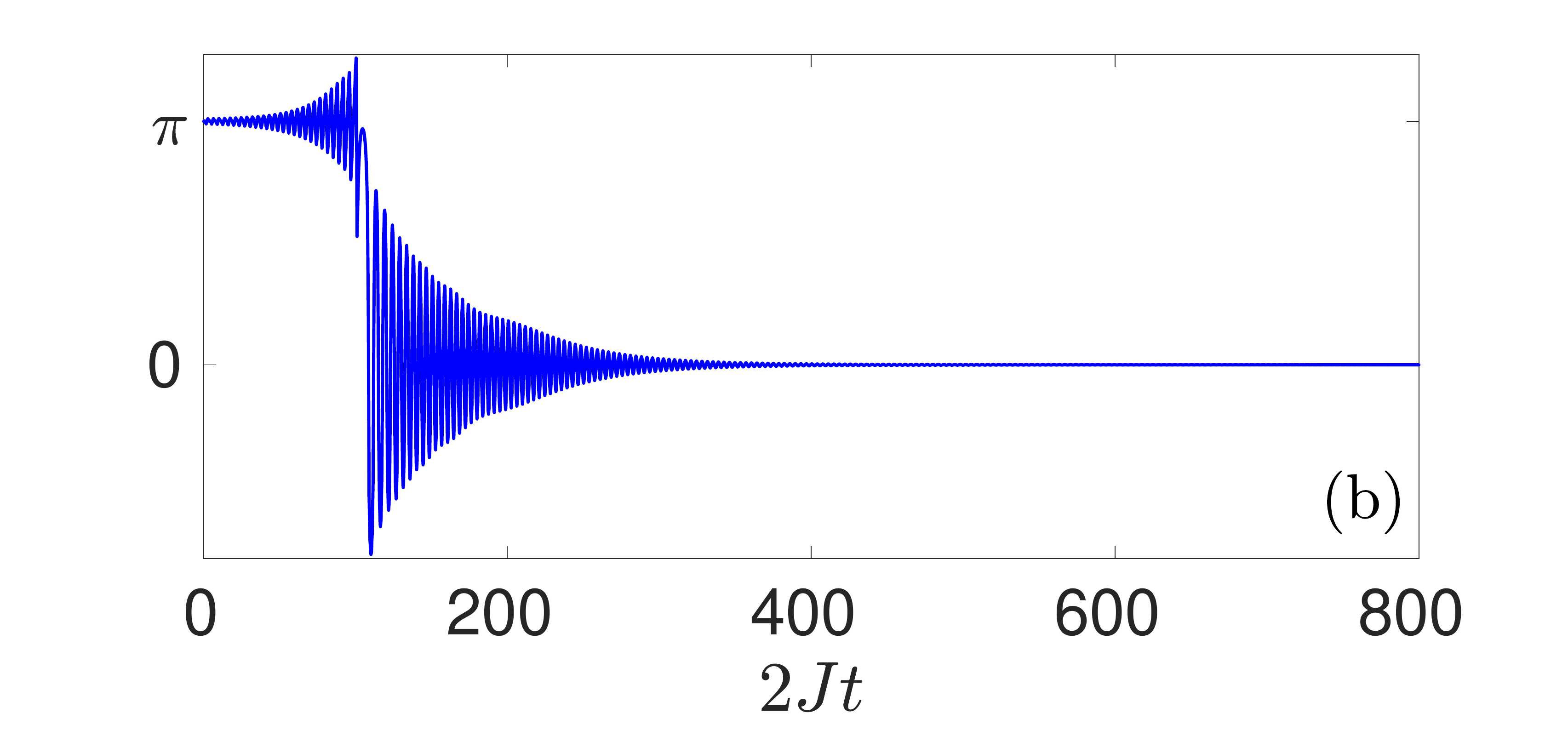}
    \end{minipage}
        \begin{minipage}{0.32\textwidth}
        \centering       
        \includegraphics[width=2.26in, height=1.8in]{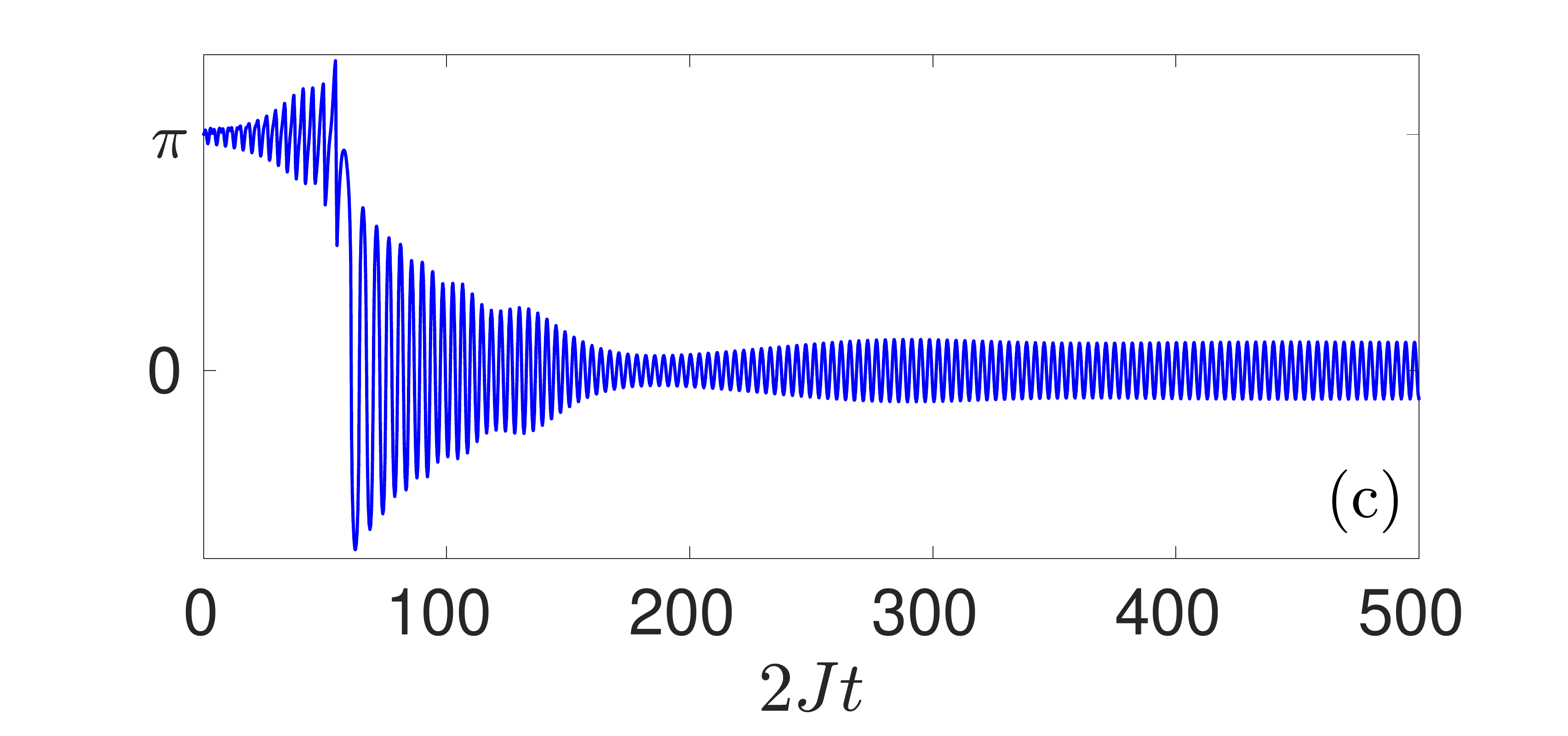}
    \end{minipage}
    \begin{minipage}{0.32\textwidth}
        \centering  
        \includegraphics[width=2.26in, height=1.8in]{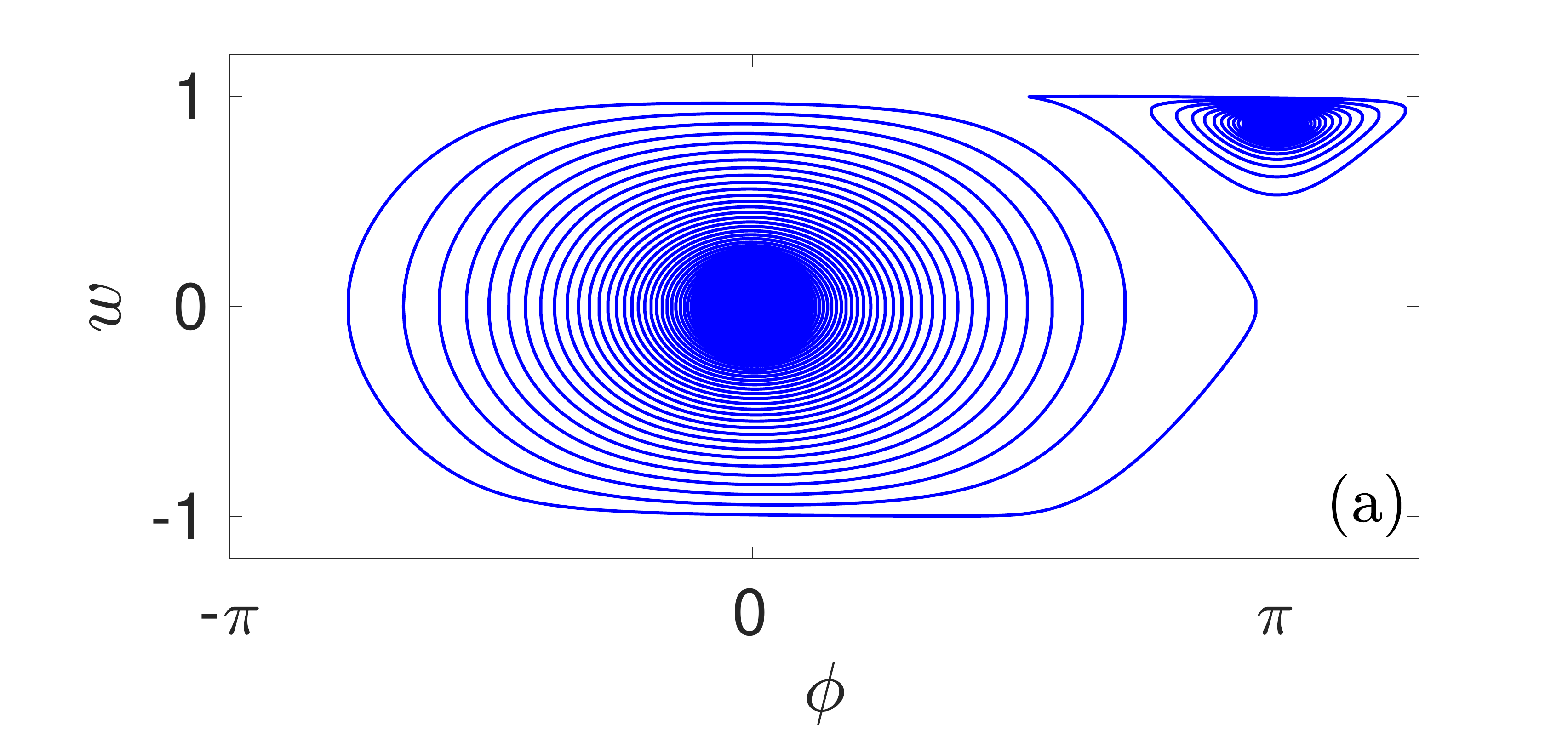}
    \end{minipage}
    \begin{minipage}{0.32\textwidth}
        \centering        
        \includegraphics[width=2.26in, height=1.8in]{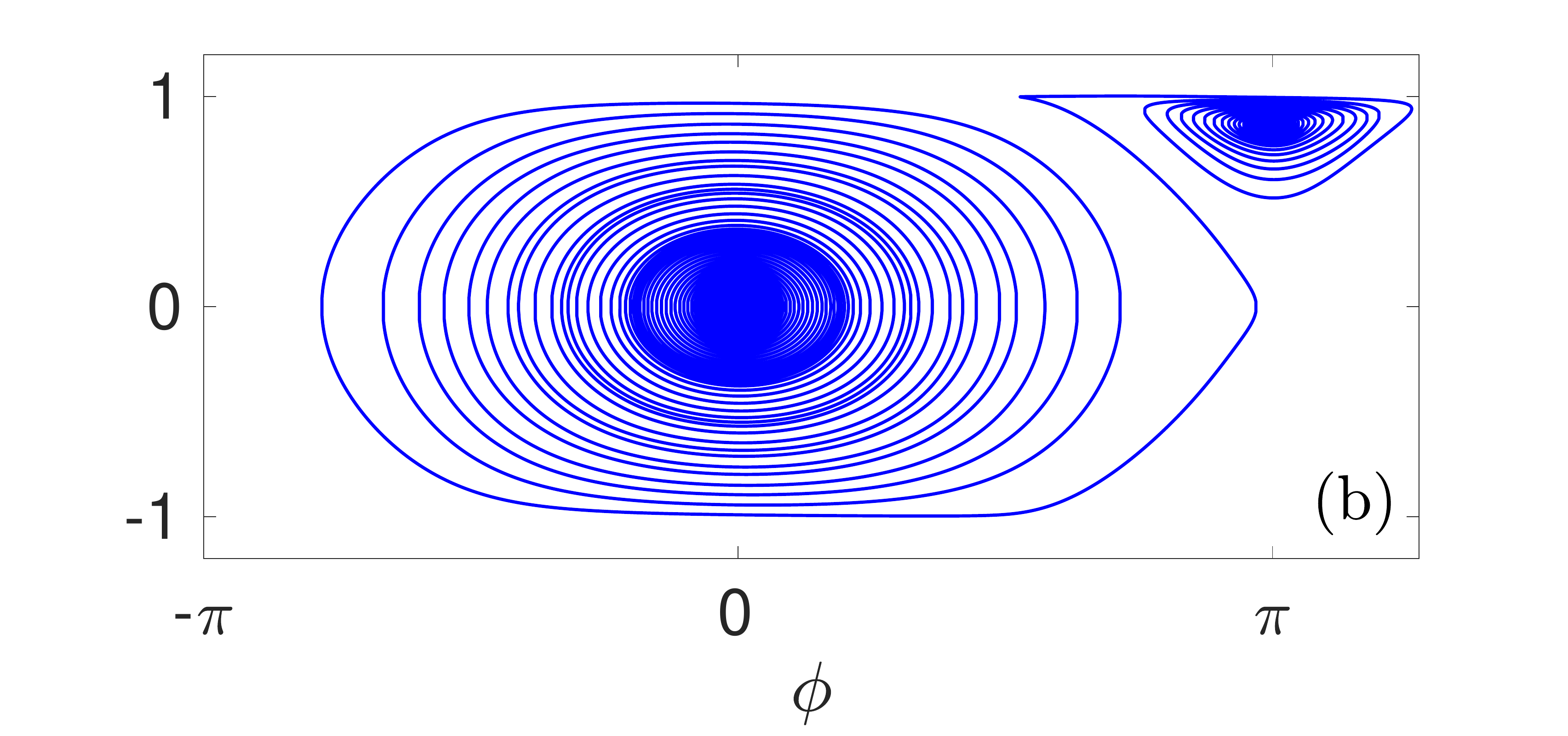}
    \end{minipage}
        \begin{minipage}{0.32\textwidth}
        \centering       
        \includegraphics[width=2.26in, height=1.8in]{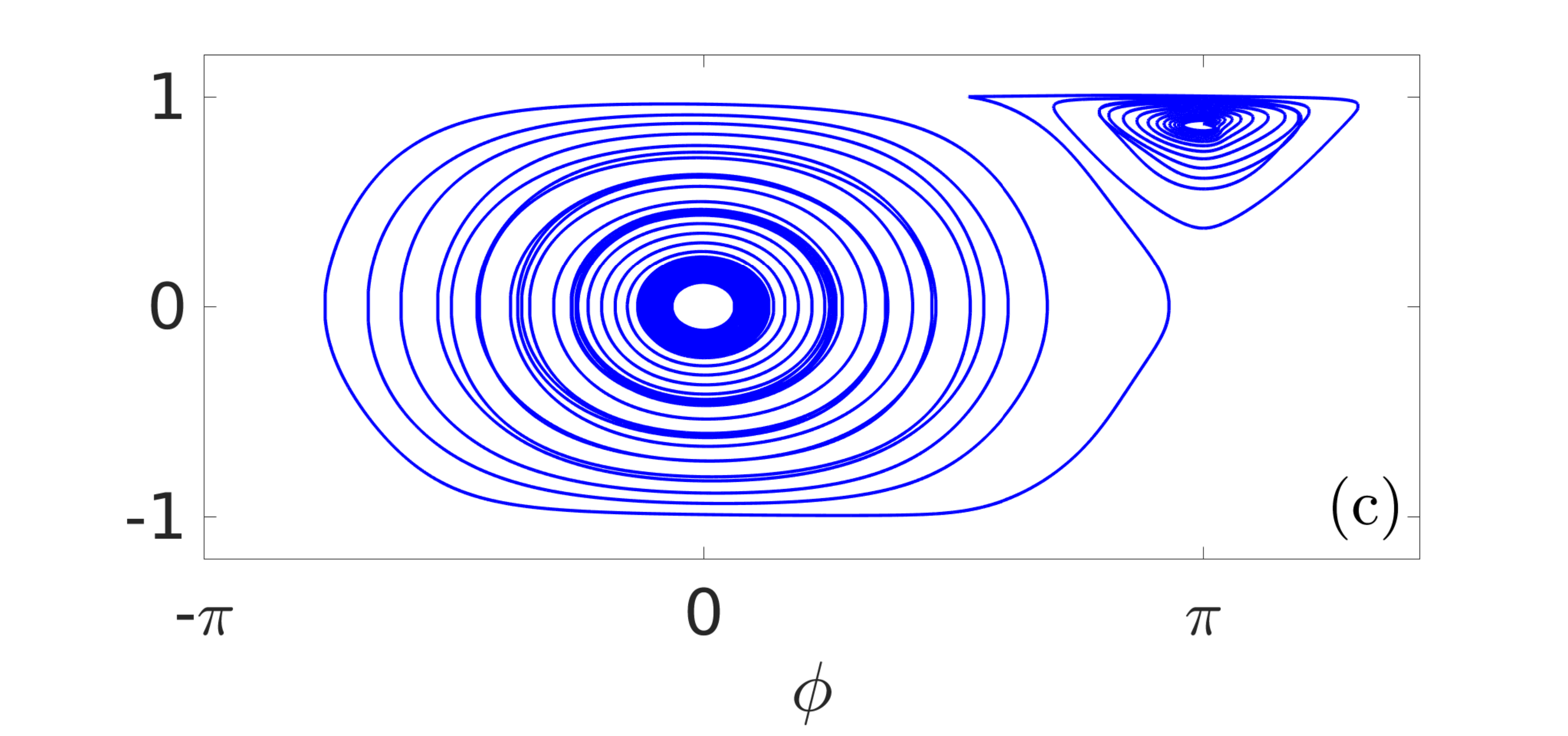}
    \end{minipage}
 \caption{\small Variation of $w(t)$ (first panel), $\phi(t)$ (second panel) as a function of $2Jt$ and phase-space trajectory (third panel) for (a) $h=0$ (b) $h=0.05$ and (c) $h=0.18$ with $w(0)=0.87$, $\phi(0)=\pi$, $\frac{\eta}{N}=0.01$, $\omega_{p}=10.20$, $NU_{0}=0.1\hbar\omega_{z}$ with $N=2000$, $J=0.024\hbar\omega_{z}$, $\zeta=0.05$ and $\kappa=0.03$  in $\pi$-phase mode.}
 \label{Fig5}
\end{figure}

To obtain MQST in $\pi$-phase mode, we assign a definite value of $\omega_{p}=3.464$ and change the amplitude of the perturbation term just below and above of the threshold value. The first panel of Fig. \ref{Fig5} shows the variation of $w(t)$ as a function of $2Jt$ for different perturbation amplitudes. It exhibits behaviour similar to that in Fig. \ref{Fig4}; the only difference is that the average of the population imbalance is non-zero which clearly describes MQST in $\pi$-phase mode and the frequency of this self-trapped state is governed by Eq. (24). When $h>h_{t}$ $(h_{t}=0.10)$, $w(t)$ starts oscillating with increasing amplitude before it jumps to the parametric oscillatory regime. So, by changing the perturbation amplitude above the threshold value, one can transform the dissipative state into non-dissipative one. In the second panel of Fig. \ref{Fig5}, we plot the time evolution of the phase difference for different perturbation amplitudes. When there is no perturbation term $\phi(t)$ oscillates around $\pi$ with increase in amplitude and after a sudden jump, it starts decaying to reach equilibrium. However, this decay in the $\pi$-phase mode around zero state can be completely nullified or mitigated by increasing the perturbation term above threshold value. The third panel of Fig. \ref{Fig5} shows the phase-space trajectory for the mentioned perturbation amplitudes. The trajectory spirals outward with increase in amplitude and goes through a phase slip and then damps down with decrease in amplitude. After crossing the threshold value, it spirals outward with increasing amplitude, goes through a phase slip and then execute sustained oscillation. Note that in the $\pi$-phase mode the system finally reaches the stable zero-state for equilibrium. So, the threshold condition or the sustained oscillations conditions holds only for the value, when we analytically put $\kappa=\frac{\eta}{N}[\Lambda_{0}+1]$ in Eq. (16).

\subsection{Analysis of stability boundaries, Arnold's tongue and chaotic oscillations}\label{56}
\begin{figure}[ht]
\centering
\centering
\begin{minipage}{0.48\textwidth}
\centering
\includegraphics[width=3.3in, height=2.2in]{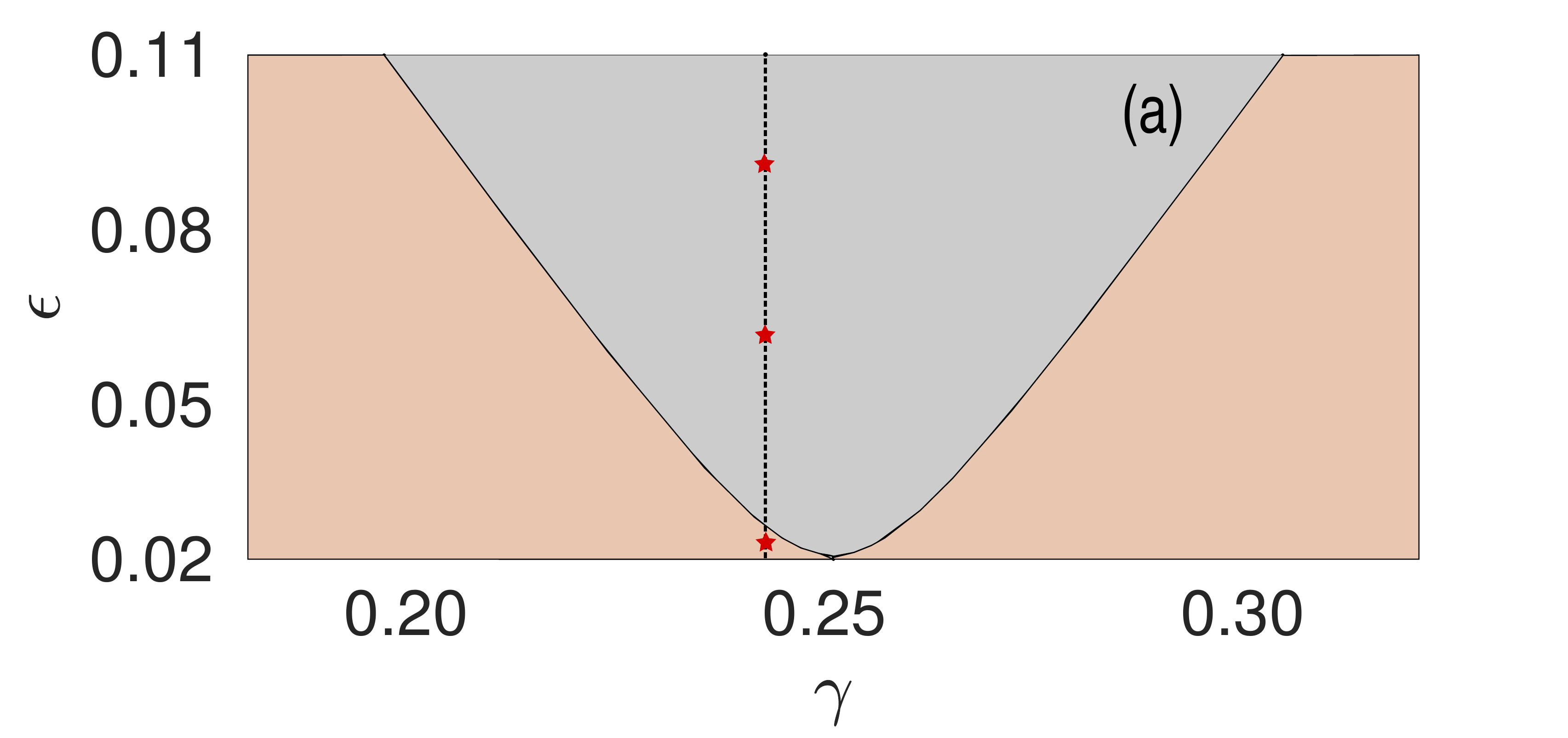}
\end{minipage}
\begin{minipage}{0.48\textwidth}
\centering
\includegraphics[width=3.3in, height=2.2in]{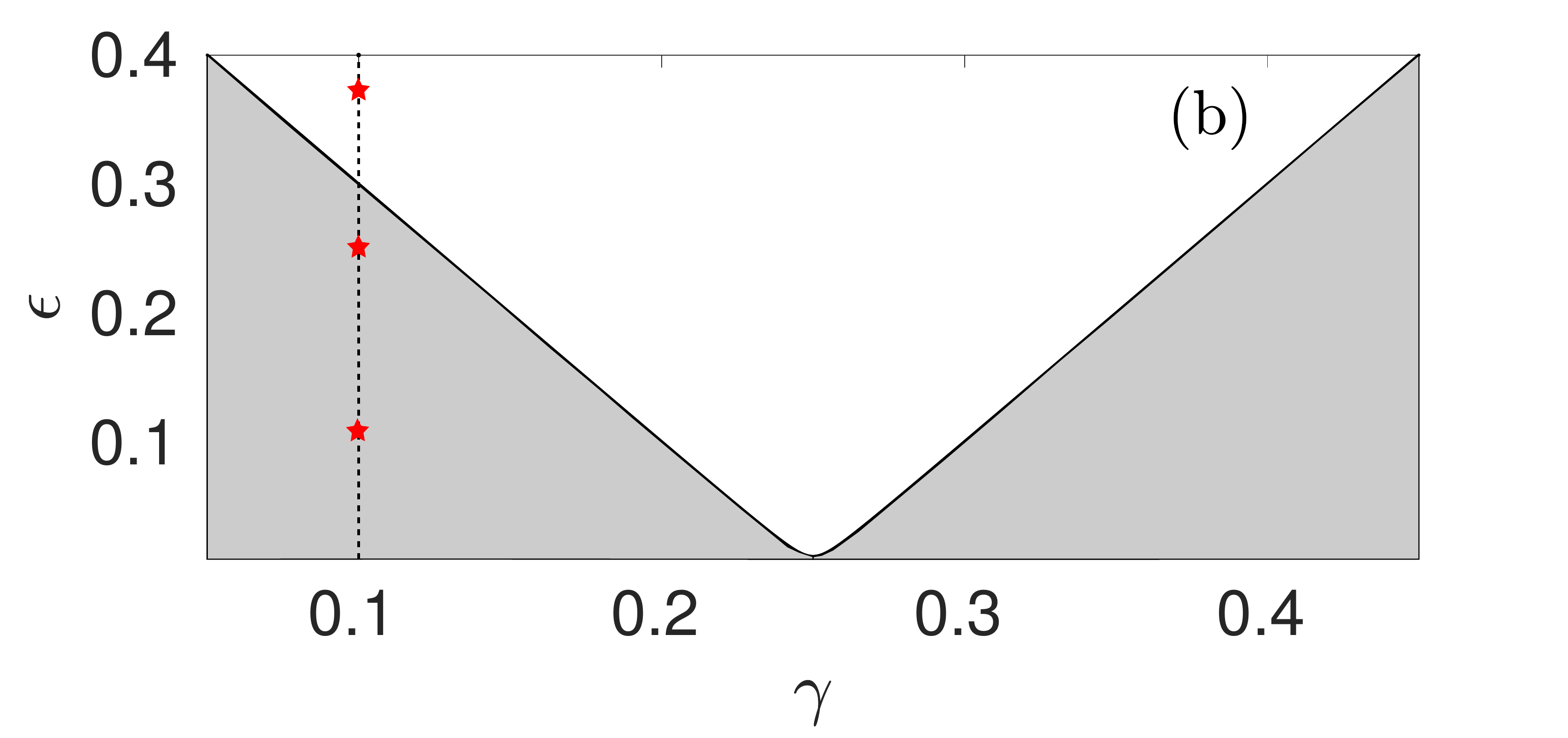}
\end{minipage}
\caption{\small Arnold's tongue for dissipative BJJ for (a) zero-phase mode and (b) $\pi$-phase mode. The three red star points in the vertical lines correspond to three different values of perturbation amplitude $h$ for fixed $\omega_{p}$, $\omega_{J}$, and $\gamma$ (see Sec. \ref{5}(\ref{56}) in the text). }
\label{Fig6}
\end{figure}

For a detailed analysis of the stable and unstable regions, we now plot $\epsilon$ vs. $\gamma$ according to Eq. (36) for the parameter values corresponding to the zero-phase mode and $\pi$-phase mode as shown in Fig. \ref{Fig6}. In zero-phase mode, we choose the value $NU_{0}=0.24\hbar\omega_{z}$, $J=0.024\hbar\omega_{z}$, $\Lambda_{0}=5$, $\omega_{p}=4.95$, $\frac{\eta}{N}=0.02$, $\kappa=0.12$. A typical parabola-shaped region in the form of well known Arnold's tongue, which separates out the parametric oscillatory regime from the steady state regime is observed. However, in the zero-phase mode, there is no unstable region for $\Lambda_{0}=5$. To get clear idea, we fix the value of $\gamma$ by setting the Josephson frequency $\omega_{J}$ and perturbation frequency $\omega_{p}$ as shown in Fig. \ref{Fig6}(a) by black dashed vertical line. Now, from $\epsilon=\frac{h}{\omega_{p}^2}$, we choose three different points (red star marked) corresponding different values of $h$ keeping $\omega_{p}$ fixed. On the dashed vertical line $\gamma=0.245$, we observe that when $\epsilon=0.021$ the perturbation amplitude smaller than the threshold value, the point is located in light brown coloured region, i.e, the steady state region. Now, when $\epsilon=0.055$ or $0.088$, the corresponding values of $h$ are greater than the threshold value, one arrives at the region with light grey colour. This corresponds to the parametric oscillatory regime.

\begin{figure}[ht]
\centering
\begin{minipage}{0.48\textwidth}
\centering
\includegraphics[width=3.3in, height=2.2in]{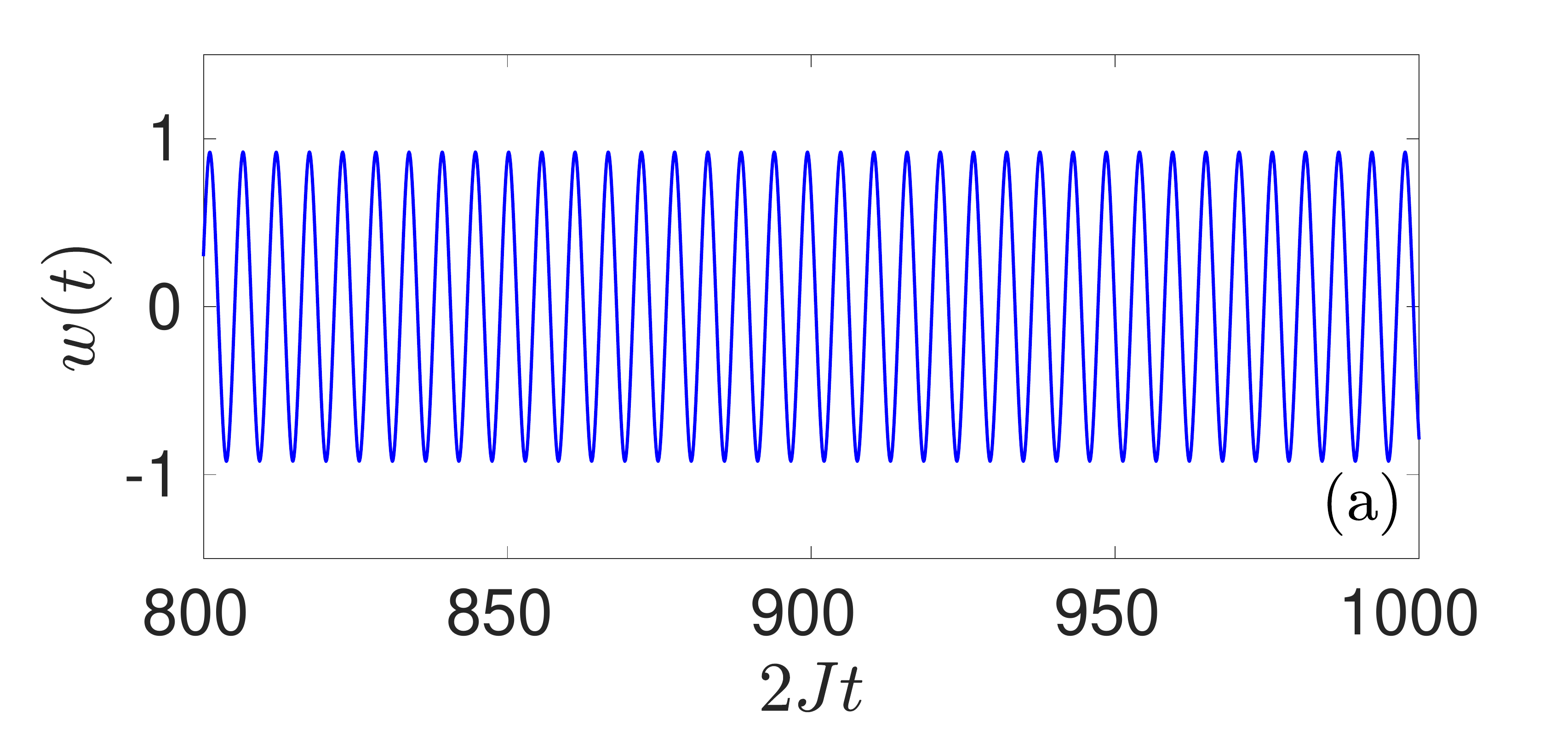}
\end{minipage}
\begin{minipage}{0.48\textwidth}
\centering
\includegraphics[width=3.3in, height=2.2in]{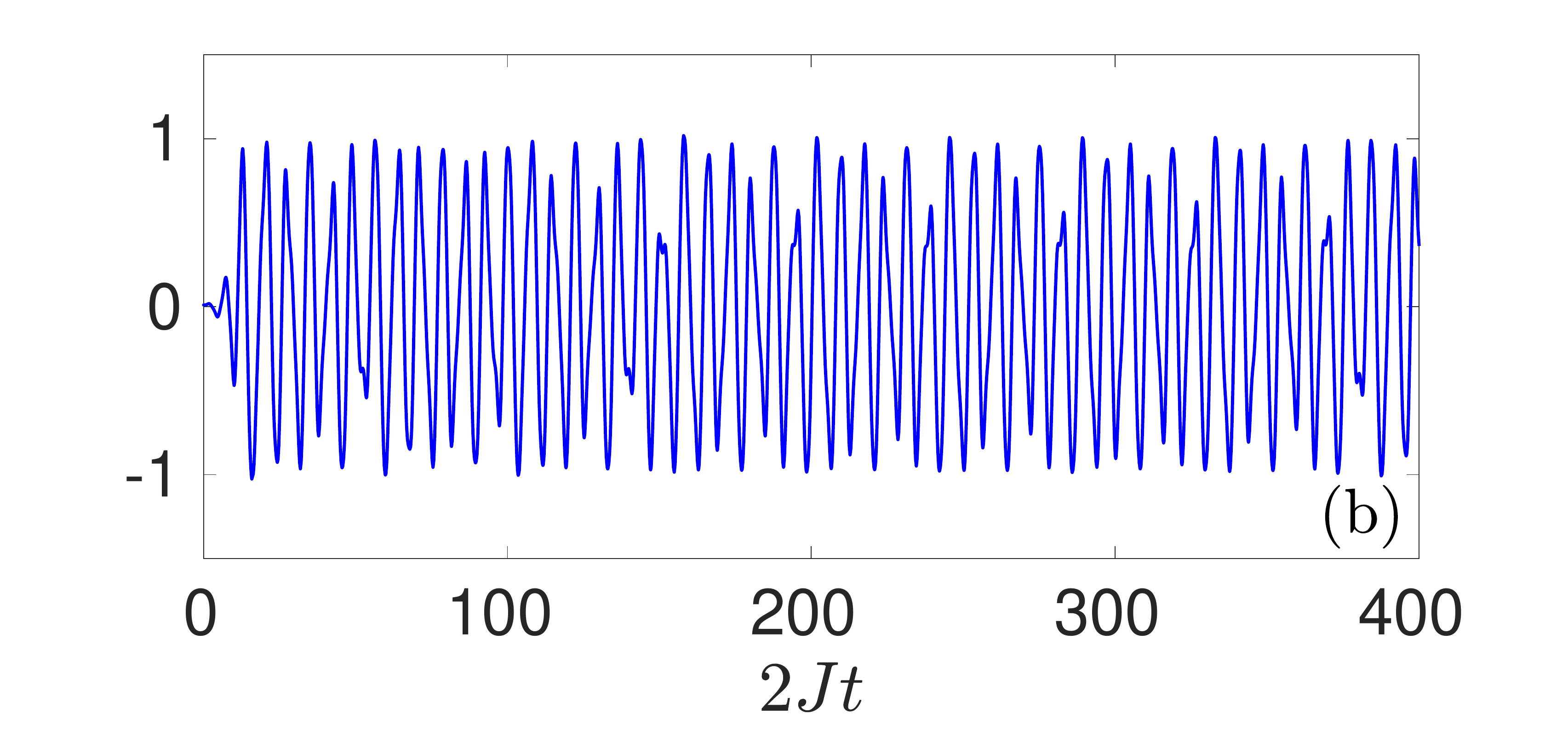}
\end{minipage}
\caption{\small Parametric (left) and chaotic (right) oscillations in $\pi$-phase mode for (a) $\epsilon=0.24$ (b) $\epsilon=0.37$.}
\label{Fig7}
\end{figure}

Now to locate the stability boundary for the $\pi$-phase mode, we choose the parameter values $NU_{0}=0.017\hbar\omega_{z}$, $J=0.024\hbar\omega_{z}$, $\Lambda_{0}=0.36$, $\omega_{p}=2.30$, $\frac{\eta}{N}=0.02$, $\kappa=0.027$. The $\epsilon$ vs. $\gamma$ plot according to Eq. (36) exhibits a V-shaped region which separates out the unstable regime from the stable parametric oscillatory regime. Similar to the zero-phase mode, here we also fix the value of $\gamma$ by setting $\omega_{J}$ and $\omega_{p}$ specific values as shown in Fig. \ref{Fig6}(b) by black dashed vertical line. From $\epsilon=\frac{h}{\omega_{p}^2}$, we choose three different points (red star marked) corresponding different values of $h$ keeping $\omega_{p}$ fixed. Now on the dashed vertical line $\gamma=0.12$, when $\epsilon=0.11$, the perturbation amplitude is higher than the threshold value, the system settles down in the parametric oscillatory regime as shown by light grey color. Further increase of $\epsilon$ to $\epsilon=0.24$, its remains still in the parametric oscillatory regime as shown in Fig. \ref{Fig7}(a)  and at $\epsilon=0.37$, the oscillations becomes aperiodic or deterministically chaotic in nature as shown in Fig. \ref{Fig7}(b). Thus the upper portion of the V-shaped regime with white colour corresponds to an unstable regime. 

One pertinent point to emphasize here is that the transition from stable steady state to periodic parametric oscillation is distinct from the transition from parametric oscillatory regime to the chaotic oscillatory regime as depicted in the Arnold's tongue  in the zero-phase mode and $\pi$-phase mode, respectively. In the zero-phase mode the boundary clearly describes the critical threshold and the stability boundary separates out the two regions, whereas in the $\pi$-phase mode the boundary demarcates the two oscillatory regions of different stability.

\section{Conclusions}\label{6}
In this paper we have considered a nonlinear dissipative BJJ subjected to a time-periodic sinusoidal perturbation of the interaction parameter. It has been shown that the dynamical system undergoes sustained periodic macroscopic quantum oscillations at a frequency half of the perturbation frequency when the strength of perturbation exceeds a critical threshold . A multiple time scale analysis of this scenario clearly reveals the domains of instability within a V-shaped region in the form of Arnold's tongue in a graph of perturbation amplitude vs. perturbation frequency. The main conclusions of this study can be summarized as follows.

(i) We have shown a new kind of sustained oscillations in a perturbed dissipative macroscopic quantum system. This periodic oscillation is quite distinct from the usual forced oscillation in a dynamical system, because, it is well-known that a forced oscillator exhibits sustained characteristic oscillations in the long time limit due to the effect of forcing for an arbitrary strength of perturbation. The periodic oscillation on the other hand discussed in the present work is a result of parametric instability and arises from the effect of internal or inherent temporal perturbation term of the system, namely, the interaction parameter.

(ii) Our numerical simulations in zero-phase mode suggest that parametric resonance is indeed possible if we overcome the losses by exceeding the critical instability threshold as stated in Eq. (16). For large perturbation amplitude, the system exhibits excitations, which, however, remains outside the scope of the present treatment.

(iii) In the running-phase mode, we have shown the transition from MQST to parametric Josephson regime when the perturbation amplitude is just above the critical threshold. In $\pi$-phase mode, we observe that the dynamics of phase difference suffers a phase slip before it executes sustained periodic oscillations.

(iv) We have also carried out a multiple time scale analysis of parametric damped oscillator to identify the different stability zones in a graph of perturbation amplitude vs. perturbation frequency. Full numerical simulation of the dynamics demonstrates the transition from stable steady state to the parametric periodic oscillatory state separated by the boundary as described by the threshold condition in Eq. (16). In the $\pi$-phase mode, we have shown a transition from the regular parametric oscillatory state to the chaotic parametric oscillatory state. The chaotic parametric oscillation is a new feature of this parametric dissipative BJJ.

The parametric dissipative BJJ studied in this paper can serve as an useful tool for probing the dynamical properties of nonlinear matter waves. We believe that parametric oscillations in dissipative BJJ will be experimentally realizable in near future with currently available ultracold atom technology.

\section*{ACKNOWLEDGMENT}
Two of us (BD and DSR) thank Department of Science and Technology (DST), Government of India for support under the project SB/S2/LOP-008/2014. Partial financial support from SERB (DST) under J. C. Bose National Fellowship is also thankfully acknowledged.

\appendix*
\section{Derivation of on-site interaction energy under sinusoidal perturbation}
The on-site interaction is calculated as 
\begin{eqnarray}
U=\int|\psi({\bf r_{1}})|^2|\psi({\bf r_{2}})|^2V_{int}({\bf r_{1}},{\bf r_{2}})d{\bf r_{1}}d{\bf r_{2}}\nonumber
\end{eqnarray}
where $\psi({\bf r})$ is the single particle 3D wave function. We consider our interatomic interaction to be of contact type, $V_{int}({\bf r_{1}},{\bf r_{2}})=\frac{4\pi\hbar^2 a_{s}}{m}\delta({\bf r_{1}}-{\bf r_{2}})$, where $a_{s}$ is the $s$-wave scattering length, $m$ is the atomic mass. As a result, $U$ becomes
\begin{eqnarray}
U=\frac{4\pi\hbar^2a_{s}}{m}\int|\psi({\bf r})|^4d{\bf r}
\end{eqnarray}
Now under harmonic approximation around the two minima of $V_{dw}$ of Eq. (5); i.e., for $z=\pm b$, 1D harmonic frequency along $z$ is $\omega_{z}(t)=\frac{2b}{\sqrt{m}}\sqrt{\chi_{0}^2+\chi_{1}^2\sin\omega_{p}t}$ which is time-dependent due to the sinusoidal modulation of the barrier height. Recent experimental and theoretical works \cite{double1,osc:1,quasi:1} have shown that the barrier height of a optical DW trap can be dynamically controlled by the laser intensity and the relative phase between the lasers and radio frequency or microwave fields in case of a magnetic DW trap \cite{double2,Schmiedmayer1,Schmiedmayer2}.

We assume that $\omega_{p}<<\omega_{z}$ and $\chi_{1}<<\chi_{0}$. If the barrier height is very large compared to the ground-state energy of a single well under harmonic approximation, then atoms will primarily occupy the lowest energy band of the DW, and the temporal modulation of the barrier height will hardly excite the system. The form of the single particle wave function is $\psi({\bf r},t)=\frac{1}{\sqrt{\pi a_{\rho}^2}}e^{-\frac{\rho^2}{2a_{\rho}^2}}\psi_{1D}(z,t)$, where $a_{\rho}=\sqrt{\frac{\hbar}{m\omega_{\rho}}}$ is the length scale in radial direction and $\omega_{\rho}$ is the radial frequency of the trap. Here $\psi_{1D}(z,t)=\frac{1}{\sqrt{\pi a_{z}(t)}}e^{-\frac{z^2}{2a_{z}^{2}(t)}}$, where $a_{z}=\sqrt{\frac{\hbar}{m\omega_{z}(t)}}$. After integrating over the radial part of Eq. (A.1), the on-site interaction becomes
\begin{eqnarray}
U=\frac{2\hbar^2a_{s}}{m a_{\rho}^2}\int|\psi_{1D}(z,t)|^4dz
\end{eqnarray}
To the first order in $\chi_{1}^2/\chi_{0}^2$, $U$ takes the form
\begin{eqnarray}
 U=U_{0}(1+\zeta\sin\omega_{p}t)
\end{eqnarray}
where $U_{0}=\frac{2a_{s}\hbar^{3/2}\sqrt{b\chi_{0}}}{\pi^{3/2}a_{\rho}^{2} m^{3/4}}$ is the unperturbed part of the on-site interaction energy and $\zeta=\frac{\chi_{1}^2}{4\chi_{0}^2}$.

\end{document}